\documentclass{article} 
\usepackage{arxiv} 
\usepackage{amsmath, amssymb, amsthm}
\usepackage{mathtools}
\usepackage{url}
\usepackage{enumitem}
\usepackage{setspace}
\usepackage{scalefnt}
\usepackage[utf8]{inputenc}
\usepackage[english]{babel}
\usepackage{bbm}
\usepackage{dsfont}
\usepackage{array}
\usepackage{blkarray, bigstrut}
\usepackage{caption} 
\usepackage{outlines}
\usepackage{enumitem}
\usepackage{hyperref} 

\usepackage{xcolor}
\usepackage{algorithm}
\PassOptionsToPackage{noend}{algpseudocode}
\usepackage{algpseudocode}
\newtheorem{theorem}{Theorem}
\newtheorem{definition}{Definition}
\newtheorem{proposition}{Proposition}

\newtheorem{corollary}{Corollary}
\newtheorem{remark}{Remark}

\DeclareMathOperator*{\argmin}{arg\,min}

\DeclarePairedDelimiter\floor{\lfloor}{\rfloor}

\setenumerate[1]{label=\arabic*}
\setenumerate[2]{label*=.\arabic*}
\setenumerate[3]{label*=.\arabic*}
\setenumerate[4]{label*=.\arabic*}

\usepackage{setspace}
\title{\vspace{-0.0em} Move Schedules: Fast persistence \\ computations  in coarse dynamic settings\thanks{This work was partially supported by the National Science Foundation through grants CCF-2006661 and CAREER award DMS-1943758.}\vspace{-0.5em}}
\author{Matt Piekenbrock\thanks{Khoury College of Computer Sciences, Northeastern University.}\;\; and Jose A. Perea$^\dagger$\thanks{Department of Mathematics and Khoury College of Computer Sciences, Northeastern University.}}
\date{}

\begin{document}
\makeatletter
\newcommand{\customlabel}[2]{%
   \protected@write \@auxout {}{\string \newlabel {#1}{{#2}{\thepage}{#2}{#1}{}} }%
   \hypertarget{#1}{#2}
}
\makeatother

\newcommand\topstrut[1][1.0ex]{\setlength\bigstrutjot{#1}{\bigstrut[t]}}
\newcommand\botstrut[1][0.9ex]{\setlength\bigstrutjot{#1}{\bigstrut[b]}}

\newcommand\undermat[2]{%
  	\makebox[0pt][l]{$\smash{\underbrace{\phantom{%
    \begin{matrix}#2\end{matrix}}}_{\text{$#1$}}}$}#2}

\maketitle  

\begin{abstract} 
Matrix reduction is the standard procedure for computing the persistent homology of a filtered simplicial complex with $m$ simplices. Its output is a particular decomposition of the total boundary matrix, from which the persistence diagrams and generating cycles are derived. 
	Persistence diagrams are known to vary continuously with respect to their input, motivating the  study of their computation for time-varying filtered complexes. Computing persistence dynamically can be reduced to maintaining a valid decomposition under adjacent transpositions in the filtration order. 
	Since there are $O(m^2)$ such transpositions, this maintenance procedure exhibits limited scalability and is often too fine for many applications. 
We propose a coarser strategy for maintaining the decomposition over a 1-parameter family of filtrations. By reduction to a particular longest common subsequence problem, we show that the minimal number of decomposition updates $d$ can be found in $O(m \log \log m)$ time and $O(m)$ space, and that the corresponding sequence of permutations---which we call a \emph{schedule}---can be constructed in $O(d m \log m)$ time.  
We also show that, in expectation, the storage needed to employ this strategy is actually sublinear in $m$. 
Exploiting this connection, we show experimentally that the decrease in operations to compute diagrams across a family of filtrations is proportional to the difference between the expected quadratic number of states and the proposed sublinear coarsening.
Applications to video data, dynamic metric space data, and multiparameter persistence are also presented.
\end{abstract} 




\section{Introduction} 
Given a triangulable topological space equipped with a tame continuous function, persistent homology captures the changes in topology across the sublevel sets of the space, and encodes them in a persistence diagram. The stability of persistence contends that   if the function changes continuously, so too will the points on the persistence diagram~\cite{cohen2007stability, cohen2006vines}. 
This motivates the  application of persistence to time-varying  settings, like that of dynamic metric spaces~\cite{kim2020spatiotemporal}. 
As persistence-related computations tend to exhibit high algorithmic complexity---essentially cubic\footnote{For finite fields, it is known that the persistence computation reduces to the PLU factorization problem, which takes $O(m^\omega)$ where $\omega \approx 2.373$ is the matrix multiplication constant.} in the size of the underlying filtration~\cite{morozov2005persistence}---their adoption to dynamic settings poses a challenging computational problem.
Currently, there is no recourse when faced with a time-varying complex containing millions of simplices across thousands of snapshots in time.
Acquiring such a capability has far-reaching consequences: methods that vectorize persistence diagrams for machine learning purposes all immediately become computationally viable tools in the dynamic setting. Such persistence summaries include adaptive template functions~\cite{polanco2019adaptive}, persistence images~\cite{adams2017persistence}, and $\alpha$-smoothed Betti curves~\cite{ulmer2019topological}. 
  
Cohen-Steiner et al. refer to a continuous 1-parameter family of persistence diagrams as a \emph{vineyard}, and they give in \cite{cohen2006vines} an efficient algorithm for their computation. 
The vineyards approach can be interpreted as an extension of the \emph{reduction} algorithm~\cite{zomorodian2005computing}, which computes the persistence diagrams of a filtered simplicial complex $K$ with $m$ simplices in $O(m^3)$ time, via a particular decomposition $R = D V$ (or $RU = D$) of the boundary matrix $D$ of $K$.
The vineyards algorithm, in turn,  transforms a time-varying filtration into a certain set of permutations of the decomposition $R = DV$, each of which takes at most $O(m)$ time to execute. If one is interested in understanding how the persistent homology of a continuous function changes over time, then this algorithm is sufficient, for homological critical points can only occur when the filtration order changes. 
Moreover, the vineyards algorithm is efficient asymptotically: if there are $d$ time-points where the filtration order changes, then  \emph{vineyards}  takes $O(m^3 + md)$ time; one initial $O(m^3)$-time reduction at time $t_0$ followed by one $O(m)$ operation to update the decomposition at the remaining time points $(t_1, t_2, \dots, t_d)$. When $d >> m$, the initial reduction cost is amortized by the cost of maintaining the decomposition, implying each diagram produced takes just linear time per time point to obtain. 

Despite its theoretical efficiency, \emph{vineyards} is often not the method of choice in practical settings. 
While there is an increasingly rich ecosystem of software packages offering variations of the standard reduction algorithm (e.g. Ripser, PHAT, Dionysus, etc. see~\cite{otter2017roadmap} for an overview), implementations of the vineyards algorithm are relatively uncommon.\footnote{Dionysus 1 does have an implementation of \emph{vineyards}, however the algorithm was never ported to version 2. Other major packages, such as GUDHI and PHAT, do not have \emph{vineyards} implementations.} 
The reason for this disparity is perhaps explained by Lesnick and Wright~\cite{lesnick2015interactive}: ``While an update to an $RU$ decomposition involving few transpositions is very fast in practice... many transpositions can be quite slow... it is sometimes much faster to simply recompute the $RU$-decomposition from scratch using the standard persistence algorithm.'' Indeed, they observe that maintaining the decomposition along a certain parameterized family is the most computationally demanding aspect of RIVET~\cite{rivet}, a software for computing two-parameter persistent homology.

The work presented here seeks to further understand and remedy this discrepancy: building on the work presented in~\cite{busaryev2010tracking}, we introduce a coarser approach to the vineyards algorithm.
Though the vineyards algorithm is efficient at constructing a \emph{continuous} 1-parameter family of diagrams, it is not necessarily efficient when the parameter is coarsely discretized.
Our methodology is based on the observation that practitioners often don't need (or want!) \emph{all} the persistence diagrams generated by a continuous 1-parameter of filtrations; usually just $n << d$ of them  suffice.   
By exploiting the ``donor'' concept introduced in~\cite{busaryev2010tracking}, we are able to make a trade-off between the number of times the decomposition is restored to a valid state and the granularity of the decomposition repair step, reducing the total number of column operations needed to apply an arbitrary permutation to the filtration. This trade-off, paired with a fast greedy heuristic explained in section~\ref{sec:proxy_objective}, yields an algorithm that can update a $R = DV$ decomposition more efficiently than \emph{vineyards} in coarse time-varying contexts, making dynamic persistence more computationally tractable for a wider class of use-cases. 
Both the source code containing the algorithm we propose and the experiments performed in Section~\ref{sec:results} are available open source online.\footnote{For all accompanying software and materials, see: \url{https://github.com/peekxc/move_schedules}}
  
\subsection{Related Work}\label{sec:related_work} 
To the author's knowledge, work focused on ways of updating a  decomposition $R = DV$, for all homological dimensions, is limited: there is the \emph{vineyards} algorithm~\cite{cohen2006vines} and the \emph{moves} algorithm~\cite{busaryev2010tracking}, both of which are discussed extensively in section~\ref{sec:background}. At the time of writing, we were made aware of very recent work~\cite{luo2021accelerating} that iteratively repairs a permuted decomposition via a column swapping strategy, which they call ``warm starts.'' Though their motivation is similar to our own, their approach relies on the reduction algorithm as a subprocedure, which is quite different from the strategy we employ here.

Contrasting the dynamic setting, there is extensive work  on improving the efficiency of computing a single (static) $R = DV$ decomposition. Chen~\cite{chen2011persistent} proposed \emph{persistence with a twist}, also called the \emph{clearing optimization}, which exploits a boundary/cycle relationship to ``kill'' columns early in the reduction rather than reducing them. 
Another popular optimization is to utilize the duality between homology and cohomology~\cite{de2011dualities}, which dramatically improves the effectiveness of the clearing optimization~\cite{bauer2021ripser}. 
There are many other optimizations on the implementation side: the use of ranking functions defined on the combinatorial number system enables implicit cofacet enumeration, removing the need to store the boundary matrix explicitly; the apparent/emergent pairs optimization identifies columns whose pivot entries are unaffected by the reduction algorithm, reducing the total number of columns which need to be reduced; sparse data structures such as bit-trees and lazy heaps allow for efficient column-wise additions with $\mathbb{Z}_2 = \mathbb{Z}/2\mathbb{Z}$ coefficients and effective $O(1)$ pivot entry retrieval, and so on~\cite{bauer2021ripser, bauer2017phat}. 

By making stronger assumptions on the underlying topological space, restricting the homological dimension, or targeting  a weaker invariant (e.g. Betti numbers), one can usually obtain faster algorithms.
For example, Attali et al.~\cite{attali2009persistence} give a linear time algorithm for computing persistence on graphs.
In the same paper, they describe how to obtain $\epsilon$-simplifications of $1$-dimensional persistence diagrams for filtered $2$-manifolds by using duality and symmetry theorems. 
Along a similar vein, Edelsbrunner et al.~\cite{edelsbrunner2000topological} give a fast incremental algorithm for computing persistent Betti numbers up to dimension $2$, again by utilizing symmetry, duality, and ``time-reversal''~\cite{delfinado1995incremental}. Chen \& Kerber~\cite{chen2013output} give an output-sensitive method for computing persistent homology,  utilizing the property that certain submatrices of  $D$ have the same rank as  $R$, which they exploit through fast sub-cubic rank algorithms specialized for sparse-matrices.  

If  zeroth homology  is the only dimension of interest, computing and updating both the persistence and rank information  is greatly simplified. For example, if the edges of the graph are in filtered order a priori, obtaining a tree representation fully characterizing the connectivity of the underlying space (also known as the \emph{incremental connectivity} problem) takes just $O(\alpha(n) n)$ time using the disjoint-set data structure, where $\alpha(n)$ is the extremely slow-growing inverse Ackermann function. 
Adapting this approach to the time-varying setting, Oesterling et al.~\cite{oesterling2015computing} give an algorithm that maintains a \emph{merge tree} with $e$ edges in $O(e)$ time per-update.
If only Betti numbers are needed, the zeroth-dimension problem reduces even further to the \emph{dynamic connectivity problem}, which can be efficiently solved in amortized $O(\log n)$ query and update times using either Link-cut trees or multi-level Euler tour trees~\cite{kapron2013dynamic}.
  
\subsection{A Motivating Example}\label{sec:motivation} 
\begin{figure}[!htb]
    \centering
    \includegraphics[height=1.25in]{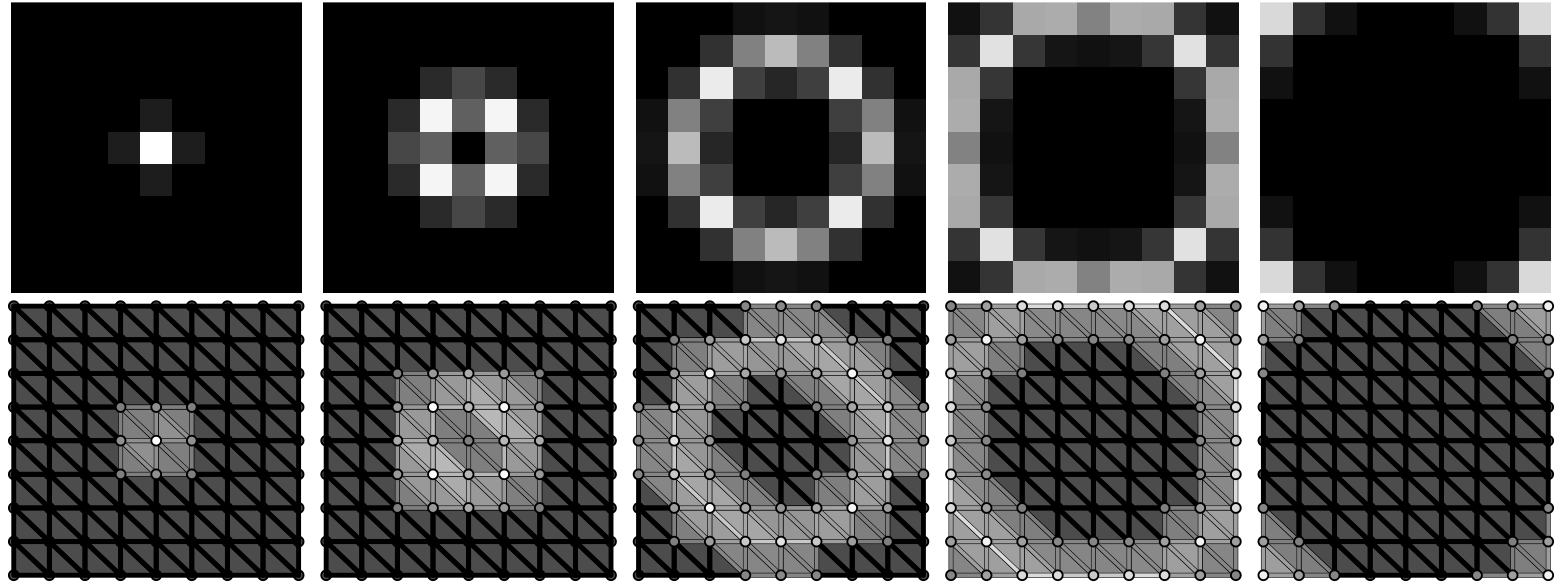}
    \caption{Top: A video of an expanding annulus. Bottom: Sublevel-set filtrations, via negative pixel intensity, of a Freudenthal triangulation of the plane.}
    \label{fig:vidExample}
\end{figure}

To motivate this effort, we begin with an illustrative example of why the vineyards algorithm does not always yield an efficient strategy for time-varying settings. 
Consider a series of grayscale images (i.e. a video) depicting a fixed-width annulus expanding about the center of a $9 \times 9$ grid, and its associated sublevel-set filtrations, as shown in Figure~\ref{fig:vidExample}.   

Each image in the series consists of pixels whose intensities vary with time, upon which we build a simplicial complex using the \emph{Freudenthal} triangulation of the plane. For each complex, we create a filtration of simplices whose order is determined by the lower stars of pixel values.   
Two events critically change the persistence diagrams: the first occurs when the central connected component splits to form a cycle, and the second  when the annulus  splits into four components. 
From left to right, the $\epsilon$-persistent Betti numbers\footnote{By ``$\epsilon$-persistent Betti number'', we mean the number of persistent pairs lying above the diagonal with persistence at least $\epsilon$, for some suitable choice of $\epsilon > 0$.} of the five evenly spaced `snapshots' of the filtration shown in Figure~\ref{fig:vidExample} are: $(\beta_0,\beta_1)  = (1,0), \; (1,1),  \; (1,1), \; (1,1), \; (4,0)$.
Thus, in this example, only a few persistence diagrams are needed to capture the major changes to the topology. 

We use this data set as a baseline for comparing \emph{vineyards} and the standard reduction algorithm $\texttt{pHcol}$ (Algorithm \ref{alg:reduce}). Suppose a practitioner wanted to know the major homological changes a time-varying filtration encounters over time.
Since it is unknown a priori when the persistent pairing function changes, one solution is to do $n$ independent persistence computations at $n$ evenly spaced points in the time domain.  An alternative approach is to construct a homotopy between a pair of filtrations $(K, f)$, $(K,f')$ and then decompose this homotopy into adjacent transpositions based on the filtration order---the \emph{vineyards} approach. 
 We refer to the former as the \emph{discrete setting}, which is often used in practice, and the  latter as the \emph{continuous setting}. Note that though the discrete setting is often more practical, it is not guaranteed to capture all homological changes in persistence that occur in the continuous 1-parameter family of diagrams.  

\begin{figure*}[!htb]
 	\centering
	\includegraphics[width=0.70\textwidth]{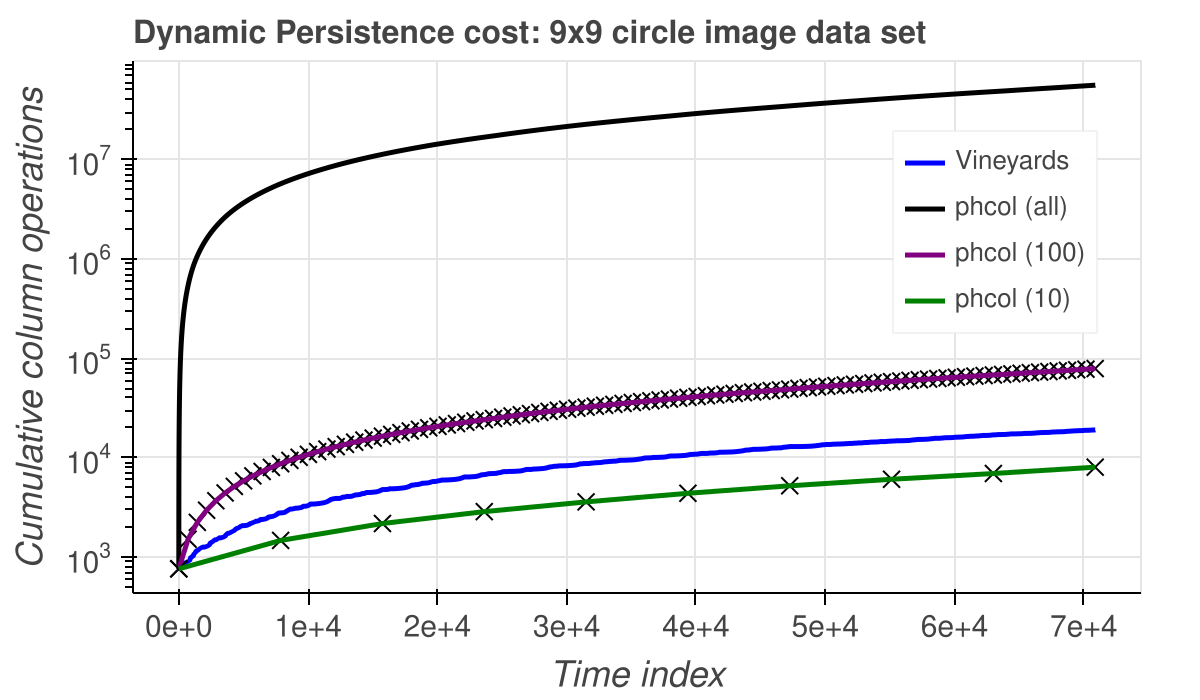}
 	\caption{The cumulative column operations needed to compute persistence across the time-varying filtration of grayscale images. Observe 10 independent persistence computations evenly spaced in time (green line) captures the major topological changes and is the most computationally efficient approach shown.}
 	\label{fig:vineyards}
 \end{figure*}

The cumulative cost (in total column operations) of these various approaches are shown in Figure~\ref{fig:vineyards}, wherein the reduction (\texttt{pHcol}) and vineyard algorithms are compared. Two discrete strategies (green and purple) and two continuous strategies (black and blue) are shown.

Note that without knowing where the persistence pairing function changes, a continuous strategy must construct all $\approx 7 \times 10^{4}$ diagrams induced by the homotopy. 
In this setting, as shown in the figure, the vineyards approach is indeed far more efficient than naively applying the reduction algorithm independently at all time points.
However, when the discretization of the time domain is coarse enough, the naive approach actually performs less column operations than \emph{vineyards}, while still capturing the main events. 
  
The existence of a time discretization that is more efficient to compute than continually updating the decomposition indicates that the vineyards framework must incur some overhead (in terms of column operations) to maintain the underlying decomposition, even when the pairing function determining the persistence diagram is unchanged. 
Indeed, as shown by the case where $n = 10$, applying \texttt{pHcol} independently between  relatively ``close'' filtrations is substantially more efficient than iteratively updating the decomposition.
 Moreover, any optimizations to the reduction algorithm (e.g. clearing~\cite{chen2011persistent}) would only increase this disparity.
Since persistence has found many applications in dynamic contexts~\cite{topaz2015topological, xian2020capturing, lesnick2015interactive, kim2020spatiotemporal}, a more efficient alternative to \emph{vineyards} is clearly needed. 
 \\
 \\
 \noindent
\textbf{Our approach and contributions are as follows:} First, we leverage the \emph{moves} framework of Busaryev et al.~\cite{busaryev2010tracking} to include  coarser  operations for dynamic persistence settings. 
By a reduction to an edit distance problem, we give a   lower bound on the minimal number of moves needed to perform an arbitrary permutation to the $R = D V$ decomposition,  along with a proof of its optimality.
We also give worst-case sizes of these quantities in expectation as well as efficient algorithms for constructing these operations---both of which are derived from a reduction to the Longest Increasing Subsequence (LIS) problem. 
These operations parameterize sequences of permutations $\mathcal{S}= ( s_1, s_2, \dots, s_d )$ of minimal size $d$, which we call \emph{schedules}. 
However, not all minimal size schedules incur the same  cost 
(i.e., number of column operations).
We investigate the feasibility of choosing optimal cost schedules, and show that 
greedy-type approaches can lead to arbitrarily bad behavior. 
In light of these results, we give an alternative proxy-objective for cost minimization, provide bounds justifying its relevance to the original problem, and give an efficient $O(d^2 \log m)$ algorithm for heuristically solving this proxy minimization. 
A performance comparison with other reduction-based persistence computations is given, wherein move schedules are demonstrated to be an order of magnitude more efficient than existing approaches at calculating persistence in dynamic settings. In particular, we illustrate the effectiveness of efficient scheduling with a variety of real-world applications, including flock analysis in dynamic metric spaces and manifold detection from image data using 2D persistence computations. 
  
\subsection{Main results}
Given a simplicial complex $K$ with filtration function $f$, denote by $R = D V$ the decomposition of its corresponding boundary matrix $D$ such that $R$ is reduced and $V$ is upper-triangular (see section~\ref{sec:reduction} for details).
 If one has a pair of filtrations $(K, f)$,  $(K, f')$ of size $m = \lvert K \rvert$ and $R = D V$ has been computed for $(K,f)$, then it may be advantageous to use the information stored in ($R$, $V$) to reduce the computation of $R' = D' V'$. 
Given a permutation $P$ such that $D' = P D P^T$, such an update scheme has the form: 
$$ (\ast P \ast R \ast P^T \ast) = (P DP^T)(\ast P \ast V \ast P^T \ast) $$
where $\ast$ is substituted with elementary column operations that repair the permuted decomposition. It is known how to linearly interpolate $f \mapsto f'$ using $d \sim O(m^2)$ updates to the decomposition, where each update  requires at most two column operations~\cite{cohen2006vines}. Since each column operation takes $O(m)$, the complexity of re-indexing $f \mapsto f'$ is $O(m^3)$, which is efficient if all $d$ decompositions are needed. Otherwise, if only $(R', V')$ is needed, updating $R \mapsto R'$ using the approach from~\cite{cohen2006vines} matches the complexity of computing $R' = D' V'$ independently. 

We now summarize our main results (Theorem~\ref{prop:sparsity_result}): suppose one has a  schedule  $\mathcal{S} = \left( s_1, s_2, \dots, s_d \right)$ yielding a corresponding sequence of decompositions:
  \begin{equation}\label{eq:rv_seq}
   	R = R_0 = D_0 V_0 \overset{s_1}{\to} D_1 V_{1} \overset{s_2}{\to} \dots \overset{s_d}{\to} D_d V_{d} = R_d = R'
 \end{equation}
 where  $s_k = (i_k, j_k)$ for $k=1,\ldots, d$,  denotes a particular type of cyclic permutation (see section~\ref{sec:moves_setting}). If $i_k < j_k$ for all $s_k \in \mathcal{S}$, our first result extends~\cite{busaryev2010tracking} by showing that~\eqref{eq:rv_seq} can be computed  using $O(\nu)$ column operations, where: 
\begin{equation}
	\quad \nu = \sum\limits_{k = 1}^d \, \lvert \mathbb{I}_{k}\rvert + \lvert \mathbb{J}_{k}\rvert 
\end{equation}
The quantities $\lvert \mathbb{I}_{k} \rvert$ and $\lvert \mathbb{J}_{k} \rvert$ depend on the sparsity of the matrices $V_k$ and $R_k$, respectively, and $d \sim O(m)$ is a constant that depends on how similar $f$ and $f'$ are. As this result depends explicitly on the sparsity pattern of the decomposition itself, it is an output sensitive bound. 

Our second result turns towards lower bounding $d = \lvert \mathcal{S} \rvert$ and the complexity of constructing $\mathcal{S}$ itself. 
By reinterpreting a special set of cyclic permutations as edit operations on strings, we find that any sequence mapping $f $ to $  f'$ of minimal size must have length (Proposition~\ref{prop:schedule_size}): 
\begin{equation}
	d = m - \lvert \mathrm{LCS}(f, f') \rvert 
\end{equation}
where $\mathrm{LCS}(f,f')$ refers to the size of the longest common subsequence between the simplexwise filtrations $(K, f)$ and $(K,f')$ (see section~\ref{sec:moves_setting} for more details).
We also show that the information needed to construct any $\mathcal{S}$ with optimal size can be computed in $O(m \log \log m)$ preprocessing time and $O(m)$ memory. We provide evidence  that $d \sim m - \sqrt{m}$ in expectation for random filtrations (Corollary~\ref{cor:expectation}). Although this implies $d$ can be $O(m)$ for pathological inputs, we give empirical results suggesting $d$  can be much smaller in practice. 
\\
\\
\noindent 
\textbf{Outline:} The paper is organized as follows: we review and establish the notations we will use to describe simplicial complexes, persistent homology, and dynamic persistence in Section~\ref{sec:background}. 
We also cover the reduction algorithm (designated here as \texttt{pHcol}), the \emph{vineyards} algorithm, and the set of \emph{move}-related algorithms introduced in~\cite{busaryev2010tracking}, which serves as the starting point of this work. 
In Section \ref{sec:move_schedules} we introduce \emph{move schedules}, and provide efficient algorithms to construct them. 
In Section \ref{sec:results} we present applications of the proposed method, including applications related to dynamic metric spaces and 2-parameter persistence. 
In Section \ref{sec:conclusion} we conclude the paper by discussing other possible applications and future work.

\section{Background}\label{sec:background} 
Suppose one has a family $\{K_i\}_{i\in I}$ of simplicial complexes indexed by a totally ordered   set $I$, and so that for any $i< j \in I$ we have $K_i \subseteq K_j$. 
Such a family is called a \emph{filtration},
which is deemed \emph{simplexwise} if $K_j \smallsetminus K_i = \{\sigma_j\}$ 
whenever $j$ is the immediate successor of $i$ in $I$.
Any finite filtration may be trivially converted into a simplexwise filtration via a set of \emph{condensing}, \emph{refining}, and \emph{reindexing} maps (see~\cite{bauer2021ripser} for more details). 
Equivalently, a filtration can be also defined as a pair $(K, f)$ where $K$ is a simplicial complex and $f : K \to I$ is a \emph{filter function}  satisfying $f(\tau) \leq f(\sigma)$ in $I$, whenever $\tau \subseteq \sigma$ in  $  K$. In this setting, $K_i = \{ \, \sigma \in K :  f(\sigma) \leq i \, \}$. Here, we consider two index sets: $[m] = \{ 1, \dots, m\}$ and $\mathbb{R}$. 
Without loss of generality, we exclusively consider simplexwise filtrations, but for brevity-sake refer to them simply as filtrations. 

Let $K$ be an abstract simplicial complex and $\mathbb{F}$ a field.
A $p$-chain is a
formal $\mathbb{F}$-linear combination  of $p$-simplices of $K$. The collection of $p$-chains under addition yields an 
$\mathbb{F}$-vector space  denoted   $C_p(K)$. 
The $p$-boundary $\partial_p(\sigma)$  of a $p$-simplex $\sigma\in K$ is the alternating sum of its oriented co-dimension 1 faces,
and the $p$-boundary of a $p$-chain is defined 
linearly in terms of its constitutive simplices. 
A $p$-chain with zero boundary is called a $p$-cycle, and together they form $Z_p(K) = \mathrm{Ker}\,\partial_p$. 
Similarly, the collection of $p$-boundaries forms  $B_p(K) = \mathrm{Im}\,\partial_{p+1}$.
Since $\partial_p \circ \partial_{p+1} = 0$ for all $p\geq 0$, 
then the quotient space $H_p(K) = Z_p(K) / B_{p}(K)$ is well-defined, and called the 
$p$-th homology of $K$ with coefficients in $\mathbb{F}$. 
If $ f: K \rightarrow  [m]$ is a filtration, then the inclusion maps  $K_i\subseteq K_{i+1}$   induce linear transformations 
at the level of homology:
\begin{equation}
	H_p(K_1) \to H_p(K_2) \to \dots \to H_p(K_m)
\end{equation}
Simplices whose inclusion in the filtration creates a new homology class are called \emph{creators}, and simplices that destroy homology classes are   called \emph{destroyers}. 
The filtration indices of 
these creators/destroyers are referred to as \emph{birth} and \emph{death} times, respectively. 
The collection of birth/death  pairs 
$(i,j)$ is denoted $\mathrm{dgm}_p(K ,f )$, 
and referred to as the $p$-th \emph{persistence diagram} of $(K,f)$.
If a homology class is born at   $K_i$ and dies entering   $K_j$, the difference $\lvert i - j \rvert$ is called the \emph{persistence} of that class.
In practice, filtrations often arise from triangulations parameterized by geometric scaling parameters, and the ``persistence'' of a homology class refers to its lifetime with respect to the scaling parameter. 

Let $\mathbb{X}$ be a triangulable topological space; that is, so that there exists   an abstract simplicial complex $K$ whose geometric realization is homeomorphic to $\mathbb{X}$. 
Let  $f: \mathbb{X} \to \mathbb{R}$ be continuous and  write $\mathbb{X}_a = f^{-1}(-\infty, a]$ to denote the sublevel sets of $\mathbb{X}$ defined by the value $a$. 
A \emph{homological critical value} of $f$ is any value $a \in \mathbb{R}$ such that the homology of the sublevel sets of $f$ changes at $a$, i.e.  if for some $p$ the inclusion-induced homomorphism  $H_p(\mathbb{X}_{a - \epsilon}) \to H_p(\mathbb{X}_{a+\epsilon})$ is not an isomorphism for any small enough $\epsilon >0$. If there are only finitely many of these homological critical values, then $f$ is said to be \emph{tame}. The concept of homological critical points and tameness will be revisited in section~\ref{sec:vineyards}.



\subsection{The Reduction Algorithm}\label{sec:reduction}
In this section, we briefly recount the original reduction algorithm introduced in~\cite{zomorodian2005computing}, also sometimes called the \emph{standard} algorithm or more explicitly \texttt{pHcol}~\cite{de2011dualities}. 
The pseudocode is outlined in Algorithm~\ref{alg:reduce} in the appendix. Without optimizations, like   clearing   or implicit matrix reduction, the standard  algorithm is very inefficient. Nonetheless, it serves as the foundation of most persistent homology implementations, and its invariants are necessary before introducing both \emph{vineyards} in section~\ref{sec:vineyards} and our move schedules in section~\ref{sec:move_schedules}.

 Given a filtration $(K, f)$ with $m$ simplices,  the output of the reduction algorithm is a matrix decomposition $R = D V$, where the persistence diagrams are encoded in $R$ and the generating cycles in the columns of $V$.
 To begin the reduction, one first  assembles the elementary boundary chains $\partial(\sigma)$ as columns ordered according to $f$ into a $m \times m$  \emph{filtration boundary matrix} $D$. Setting $V = I$ and $R = D$, one proceeds by performing elementary left-to-right column operations on $V$ and $R$ until the following invariants are satisfied:
\\
\\
\textbf{ Decomposition Invariants:}
 \begin{enumerate}[labelsep=3pt, topsep=1pt,itemsep=-0.25ex,parsep=1.0ex]\label{inv:decomposition}
 	\item[I1.] $R = D V$ where $D$ is the boundary matrix of the filtration $(K,f)$
 	\item[I2.] $V$ is full-rank upper-triangular
 	\item[l3.] $R$ is \emph{reduced}: if $\mathrm{col}_i(R) \neq 0$ and $\mathrm{col}_j(R) \neq 0$, then $\mathrm{low}_R(i) \neq \mathrm{low}_R(j)$ 
 	\end{enumerate} 
 where $\mathrm{low}_R(i)$ denotes the largest row index of a non-zero entry in column $i$ of $R$. 
We call the decomposition satisfying these three invariants \emph{valid}. The persistence diagrams of the corresponding filtration  can be determined from the lowest entries in $R$. Note that though $R$ and $V$ are not unique, the collection of persistent pairings are~\cite{zomorodian2005computing}.
 
 It is at times more succinct to restrict to specific sub-matrices of $D$ based on the homology dimension $p$, and so we write $D_p$ to represent the $d_{p-1} \times d_p$ matrix representing $\partial_p$ (the same notation is extended to $R$ and $V$).
We illustrate the reduction algorithm with an example   below. 
\\
\\
\noindent
\textbf{Example 2.1:} Consider a triangle with vertices $u,v,w$, edges $a = (u,w)$, $b = (v,w)$, $c = (u, v)$,
and whose filtration order is given as $ (u, v, w, a, b, c)$. 
Using $\mathbb{Z}_2$ coefficients, the reduction proceeds to compute $(R_1,V_1)$ as follows:
$$
	\def\rowsp{0.4em}
	\def\arraystretch{0.95}
	\makeatletter\setlength\BA@colsep{1.2pt}\makeatother
	\begin{blockarray}{cccc}{D_1}
	 \kern-0.2em & a & b & c  \\
		\begin{block}{c[ccc]}
  		u \kern\rowsp  & \; 1 &    &  1 \; \topstrut \\
  		v \kern\rowsp & \;     &  1 & 1 \; \\
  		w \kern\rowsp & \; 1 &  \underline{1} &    \; \botstrut \\
		\end{block}
	\end{blockarray}
	\; , \;
	\begin{blockarray}{cccc}{I_1}
	\kern-0.2em & a & b & c  \\
		\begin{block}{c[ccc]}
        a \kern\rowsp  & \; 1 &  &  \; \topstrut \\
  		b \kern\rowsp & \;  & 1 & \; \\
  		c \kern\rowsp & \;  &  & 1 \; \botstrut \\
		\end{block}
	\end{blockarray}
	\quad \xrightarrow{} \,
	\begin{blockarray}{cccc}{}
	\kern-0.2em & a & b & c  \\
		\begin{block}{c[ccc]}
  		u \kern\rowsp  & \; 1 & 1 & 1 \; \topstrut \\
  		v \kern\rowsp & \;     & 1 & \underline{1} \; \\
  		w \kern\rowsp & \; 1 &    &    \; \botstrut \\
		\end{block}
	\end{blockarray}
	\; , \;
	\begin{blockarray}{cccc}{}
	\kern-0.2em & a & b & c  \\
		\begin{block}{c[ccc]}
  		a \kern\rowsp  & \; 1 & 1 &    \; \topstrut \\
  		b \kern\rowsp & \;     & 1 &    \; \\
  		c \kern\rowsp & \;     &    & 1 \; \botstrut \\
		\end{block}
	\end{blockarray}
	\quad \xrightarrow{} \,
	\begin{blockarray}{cccc}{R_1}
	\kern-0.2em & a & b & c  \\
		\begin{block}{c[ccc]}
  		u \kern\rowsp  & \; 1    & 1  &  \; \topstrut \\
  		v \kern\rowsp &  \;    & 1  &  \; \\
  		w \kern\rowsp & \; 1  &     &  \; \botstrut \\
		\end{block}
	\end{blockarray}
	\; , \;
	\begin{blockarray}{cccc}{V_1}
	\kern-0.2em & a & b & c  \\
		\begin{block}{c[ccc]}
  		b \kern\rowsp  & \; 1 & 1 & 1 \; \topstrut \\
  		a \kern\rowsp & \;     & 1 & 1 \; \\
  		c \kern\rowsp & \;     &    & 1 \; \botstrut \\
		\end{block}
	\end{blockarray}
$$
\noindent Since column $c$ in $R_1$ is $0$, the 1-chain  indicated by the column $c$ in $V_1$ represents a dimension $1$ cycle. Similarly, the columns at $u, v, w$ in $R_0$ (not shown) are all zero, indicating three $0$-dimensional homology classes are born, two of which are killed by the pivot entries in columns $a$ and $b$ in $R_1$.
\\
\\
\noindent
Inspection of the algorithm from~\cite{edelsbrunner2000topological} suggests an upper bound for the reduction is $O(m^3)$, where $m$ is the number of simplices of the filtration---this bound is in fact tight~\cite{morozov2005persistence}. 
Despite its high algorithmic complexity, many variations and optimizations to Algorithm~\ref{alg:reduce} have been proposed over the past decade, see~\cite{bauer2021ripser, bauer2017phat, chen2011persistent} for an overview. 
\subsection{Vineyards}\label{sec:vineyards}
\vspace{-0.5em}
Consider a homotopy $F(x,t) : \mathbb{X} \times [0,1] \to \mathbb{R}$ on a triangulable topological space $\mathbb{X}$, and denote its ``snapshot'' at a given time-point $t$ by $f_t(x) = F(x,t)$.
The snapshot $f_0$ denotes the initial function at time $t = 0$ and $f_1$ denotes the function at the last time step. 
As $t$ varies in $[0,1]$, the points in $\mathrm{dgm}_p(f_t)$ trace curves in $\mathbb{R}^3$ which, by the stability of persistence, will be continuous if $F$ is continuous and the $f_t$'s are tame.
Cohen-Steiner et al.~\cite{cohen2007stability} referred to these curves as \emph{vines}, a collection of which forms as \emph{vineyard}---the geometric analogy is meant to act as a guidepost for practitioners seeking to understand the evolution of topological structure over time.

The original purpose of \emph{vineyards}, as described in~\cite{cohen2006vines}, was to compute a continuous 1-parameter family of persistence diagrams over a time-varying filtration, detecting homological critical events along the way.
As homological critical events only occur when the filtration order changes, detecting all such events may be reduced to computing valid decompositions at   time points interleaving all changes in the filtration order. 
For simplexwise filtrations, these changes manifest as transpositions of adjacent simplices, and thus any fixed set of rules that maintains a valid $R = D V$ decomposition under adjacent column transpositions is sufficient to compute persistence dynamically. 

To ensure a decomposition is valid, these rules prescribe certain column and row operations to apply to a given matrix decomposition either before, during, or after each transposition.   
Formally, let $S_{i}^j$ represent the upper-triangular matrix such that $A S_{i}^j$ results in adding column $i$ of $A$ to column $j$ of $A$, and let $S_{i}^j A$ be the same operation on rows $i$ and $j$.
Similarly, let $P$ denote the matrix so that $A P^T$ permutes the columns of $A$ and $P A$ permutes the rows.
Since the columns of $P$ are orthonormal, $P^{-1} = P^T$, then $P A P^T$ performs the same permutation to both the columns and rows of $A$. In the special case where $P$ represents a transposition, we have $P = P^T$ and may instead simply write $P A P$. 
The goal of the vineyards algorithm can now be described explicitly: to prescribe a set of rules, written as matrices $S_{i}^{j}$, such that if $R = D V$ is a valid decomposition, then $(\ast P \ast R \ast P \ast) = (PDP)(\ast P\ast V \ast P \ast)$ is also a valid decomposition, where $\ast$ is some number (possibly zero) of matrices encoding elementary column or row operations. 
\\
\\
\noindent 
\textbf{Example 2.2} To illustrate the basic principles of \emph{vineyards}, we re-use the running example introduced in the previous section. Below, we illustrate the case of exchanging simplices $a$ and $b$ in the filtration order, and restoring $RV$ to a valid decomposition. 
$$
	\def\rowsp{0.2em}
	\def\arraystretch{0.95}
	\makeatletter\setlength\BA@colsep{1.2pt}\makeatother
	\begin{blockarray}{cccc}{R_1}
	 \kern-0.2em  \vspace{-0.3em} & a & b & c  \\
		\begin{block}{c[ccc]}
  		u \kern\rowsp  & \; 1 & 1 &  \hphantom{1} \; \topstrut \\
  		v \kern\rowsp & \;  & 1 &  \hphantom{1} \; \\
  		w \kern\rowsp & \; 1 &  &  \hphantom{1} \; \botstrut \\  
		\end{block}
	\end{blockarray}
	\xrightarrow{S_{1}^2} 
	\begin{blockarray}{cccc}{}
	\kern-0.2em \vspace{-0.3em} & a & b & c  \\
		\begin{block}{c[ccc]}
  		 & \; 1 &  &  \hphantom{1} \; \topstrut \\
  		 & \;  & 1 &  \hphantom{1} \; \\
  		 & \; 1 & 1 & \hphantom{1} \; \botstrut \\
		\end{block}
	\end{blockarray}
		\xrightarrow{P}
	\hspace{-0.2em}
	\begin{blockarray}{cccc}{}
	\kern-0.2em \vspace{-0.3em} & b & a & c  \\
		\begin{block}{c[ccc]}
  		& \;  & 1 &  \hphantom{1} \; \topstrut \\
  		& \; 1 &  &  \hphantom{1} \; \\
  		& \; 1 & \underline{1} &  \hphantom{1} \; \botstrut \\
		\end{block}
	\end{blockarray}
	\xrightarrow{S_1^2}
	\hspace{-0.1em}
	\begin{blockarray}{cccc}{}
	\kern-0.2em \vspace{-0.3em} & b & a & c  \\
		\begin{block}{c[ccc]}
  		u \kern\rowsp & \;  & 1 &  \hphantom{1} \; \topstrut \\
  		v \kern\rowsp & \; 1 & 1 &  \hphantom{1} \; \\
  		w \kern\rowsp & \; 1 &  &  \hphantom{1} \; \botstrut \\
		\end{block}
	\end{blockarray}
$$
\vspace{-1.8em}
$$
	\def\rowsp{0.2em}
	\def\arraystretch{0.95}
	\makeatletter\setlength\BA@colsep{1.2pt}\makeatother
	\begin{blockarray}{cccc}{V_1}
	\kern-0.2em \vspace{-0.3em} & a & b & c  \\
		\begin{block}{c[ccc]}
        a \kern\rowsp  & \; 1 &  \underline{1} & 1 \; \topstrut \\
  		b \kern\rowsp & \;  & 1 & 1 \; \\
  		c \kern\rowsp & \;  &  & 1 \; \botstrut \\
		\end{block}
	\end{blockarray}
	\xrightarrow{S_{1}^2} 
	\hspace{-0.1em}
	\, \,
	\begin{blockarray}{cccc}{}
	\kern-0.2em \vspace{-0.3em} & a & b & c  \\
		\begin{block}{c[ccc]}
  		 & \; 1 &  & 1 \; \topstrut \\
  		 & \;  & 1 & 1 \; \\
  		 & \;  &  & 1 \; \botstrut \\
		\end{block}
	\end{blockarray}
	\xrightarrow{P}
	\begin{blockarray}{cccc}{}
	\kern-0.2em \vspace{-0.3em} & b & a & c  \\
		\begin{block}{c[ccc]}
  		& \; 1 &  & 1 \; \topstrut \\
  		& \;  & 1 & 1 \; \\
  		& \;  &  & 1 \; \botstrut \\
		\end{block}
	\end{blockarray}
	\xrightarrow{S_1^2}
	\begin{blockarray}{cccc}{}
	\kern-0.2em \vspace{-0.3em} & b & a & c  \\
		\begin{block}{c[ccc]}
  		 b \kern\rowsp & \; 1 & 1 & 1 \; \topstrut \\
  		 a \kern\rowsp & \; & 1 & 1 \; \\
  		 c \kern\rowsp & \;  &  & 1 \; \botstrut \\
		\end{block}
	\end{blockarray}
$$
Starting with a valid reduction $R = DV$ and prior to performing the exchange, observe that the highlighted entry in $V_1$ would render $V_1$ non-upper triangular after the exchange. This entry is removed by a left-to-right column operation, given by applying  $S_1^{2}$ on the right to $R_1$ and $V_1$. After this operation, the permutation may be safely applied to $V_1$. Both before and after the permutation $P$, $R_1$ is rendered non-reduced, requiring another column operation to restore the decomposition to a valid state.

 The time complexity of \emph{vineyards} is determined entirely by the complexity of performing  adjacent transpositions.
Since column operations are the largest complexity operations needed and each column can have potentially $O(m)$ entries, the complexity of \emph{vineyards} is $O(m)$ per transposition. 
Inspection of the individual cases of the algorithm from~\cite{cohen2006vines} shows that any single transposition requires at most two such operations on both $R$ and $V$.
However, several factors can affect the runtime efficiency of the vineyards algorithm. 
On the positive side, as both the matrices $R$ and $V$ are often sparse, the cost of a given column operation is proportional to the number of non-zero  entries in the two columns being modified. Moreover, as a rule of thumb, it has been observed that most transpositions require no column operations~\cite{edelsbrunner2000topological}. 
On the negative side, one needs to frequently query the non-zero status of various entries in $R$ and $V$ (consider evaluating e.g. Case 1.1 in~\cite{cohen2006vines}), which accrues a non-trivial runtime cost due to the quadratic frequency with which they are required.   
 

\subsection{Moves}\label{sec:moves} Originally developed to accelerate tracking generators with temporal coherence, Busaryev et al.~\cite{busaryev2010tracking} introduced an extension of the vineyards algorithm which maintains a $R = D V$ decomposition under \emph{move operations}. A move operation $\mathrm{Move}(i,j)$ is a set of rules for maintaining a valid decomposition under the permutation $P$ that moves a simplex $\sigma_i$ at position $i$ to position $j$. If $j = i \pm 1$, this operation is an adjacent transposition, and in this sense \emph{moves} generalizes \emph{vineyards}.
However, the move framework presented by Busaryev is actually distinct in that it exhibits several attractive qualities not inherited by the \emph{vineyards} approach that warrants further study.  

For completeness, we recapitulate the motivation of the \emph{moves} algorithm from~\cite{busaryev2010tracking}. 
Let $f: K \rightarrow [m]$ denote a filtration of size $m = \lvert K \rvert$ and $R = DV$ its decomposition. 
Consider the permutation $P$ that moves a simplex $\sigma_i$ in $K$ to position $j$, shifting all intermediate simplices $\sigma_{i+1}, \dots, \sigma_{j}$ down by one ($i < j$). 
To perform this shift, all entries $V_{ik} \neq 0$ with column positions $k \in [i+1,j]$ need to set to zero, otherwise $PV$ is not upper-triangular. 
We may zero these entries in $V$ using column operations $V(\ast)$, ensuring invariant I2~\eqref{inv:decomposition} is maintained, however these operations may render $R' = P R(\ast) P^T$ unreduced, breaking invariant I3.
Of course, we could then reduce $R'$ with additional column operations, but the number of such operations scales $O(\lvert i - j\rvert^2)$, but this is no more efficient than simply performing the permutation and applying the reduction algorithm to columns $[i,j]$ in $R$.

To bypass this difficulty, Busaryev et al. observed that if $R$ contains $s$ pivot entries in the columns $[i,j]$, then since it is reduced, $R'$ must \emph{also} have $s$ pivots. 
Thus, if column operations render some pivot-column $r_k$ of $R$ unreduced, then its pivot entry $\mathrm{low}_R(k)$ becomes \emph{free}\footnote{The process of donating pivot columns using $O(1)$ auxiliary storage is similar in spirit to the in-place sorting algorithm \emph{cycle sort}, which is often used to sort permutations in $O(n)$ time.}---if $r_{k}$ is copied prior to modification, we may re-use or \emph{donate} its pivot entry to a later column $r_{k+1}, \dots, r_j$. 
Repeating this process at most $j - i - 1$ times ensures $R'$ stays reduced in all except possibly at its  $i$-th column. Moreover, since the $k$-th such operation simultaneously sets $v_{ik} = 0$, $V'$ retains its upper-triangularity. 
\\
\\
\noindent \textbf{Example 2.3:} We re-use the running example from sections~\ref{sec:reduction} and~\ref{sec:vineyards} to illustrate \emph{moves}. The donor columns of $R$ and $V$ are denoted as $d_R$ and $d_V$, respectively. Consider moving edge $a$ to the position of edge $c$ in the filtration. 
$$
	\def\rowsp{0.4em}
	\def\arraystretch{0.60}
	\makeatletter\setlength\BA@colsep{0.9pt}\makeatother
	\begin{blockarray}{cc}{d_R}
	\kern-0.2em & a \\
		\begin{block}{c[c]}
  		u \kern \rowsp & \; 1 \; \topstrut \\
  		v \kern \rowsp & \;  \; \\
  		w \kern \rowsp & \; 1 \; \botstrut \\
		\end{block}
	\end{blockarray}
	\; \;
	\begin{blockarray}{cccc}{R}
	 \kern-0.2em & a & b & c  \\
		\begin{block}{c[ccc]}
  		u \kern \rowsp & \; 1 & 1 &  \hphantom{1} \; \topstrut \\
  		v \kern \rowsp & \;  & 1 &  \hphantom{1} \; \\
  		w \kern \rowsp & \; 1 &  &  \hphantom{1} \; \botstrut \\
		\end{block}
	\end{blockarray}
\; \to
\begin{blockarray}{cc}{}
	\kern-0.2em & b \\
		\begin{block}{c[c]}
  		  & \; 1 \; \topstrut \\
  		 & \;  1 \; \\
  		 & \; \; \botstrut \\
		\end{block}
	\end{blockarray}
	\; \;
\begin{blockarray}{cccc}{}
	\kern-0.2em & a & b & c  \\
		\begin{block}{c[ccc]}
  		  & \; 1 &    &  \hphantom{1} \; \topstrut \\
  		  & \;    & 1 &  \hphantom{1} \; \\
  		 & \; 1 & \underline{1} &  \hphantom{1} \; \botstrut \\
		\end{block}
	\end{blockarray}
\; \to
	\begin{blockarray}{cc}{}
	\kern-0.2em & c \\
		\begin{block}{c[c]}
  		  & \; \hphantom{1} \; \topstrut \\
  		 & \;  \hphantom{1} \; \\
  		 & \; \hphantom{1} \; \botstrut \\
		\end{block}
	\end{blockarray}
	\; \; 
	\begin{blockarray}{cccc}{}
	\kern-0.2em & a & b &  c \\
		\begin{block}{c[ccc]}
  		  & \; 1   &    &   1  \; \topstrut \\
  		 & \;       &  1  &  1  \; \\
  		 & \; 1   &   \underline{1}   &     \; \botstrut \\
		\end{block}
	\end{blockarray}
\; \xrightarrow{P} 
\begin{blockarray}{cc}{}
	\kern-0.2em & a \\
		\begin{block}{c[c]}
  		  & \; \hphantom{1} \; \topstrut \\
  		 & \; \hphantom{1} \; \\
  		 & \; \hphantom{1} \; \botstrut \\
		\end{block}
	\end{blockarray}
\; \;
\begin{blockarray}{cccc}{}
	\kern-0.2em & b & c & a  \\
		\begin{block}{c[ccc]}
  		  & \;  & 1 &  1\; \topstrut \\
  		 & \; 1 & 1 &  \; \\
  		 & \; 1 &  & \underline{1} \; \botstrut \\
		\end{block}
	\end{blockarray}
\xrightarrow{d_R} 
\begin{blockarray}{cccc}
\kern-0.2em & b & c & a  \\
	\begin{block}{c[ccc]}
		& \;  & 1 &  \hphantom{1} \; \topstrut \\
		& \; 1 & 1 &  \hphantom{1} \; \\
		& \; 1 &  &  \hphantom{1} \; \botstrut \\
	\end{block}
\end{blockarray}
$$

\vspace{-2.2em}

$$
	\def\rowsp{0.4em}
	\def\arraystretch{0.60}
	\makeatletter\setlength\BA@colsep{0.9pt}\makeatother
	\begin{blockarray}{cc}{d_V}
	\kern-0.2em & a \\
		\begin{block}{c[c]}
  		a \kern \rowsp & \; 1 \; \topstrut \\
  		b \kern \rowsp & \;  \; \\
  		c \kern \rowsp & \;  \; \botstrut \\
		\end{block}
	\end{blockarray}
	\; \;
	\begin{blockarray}{cccc}{V}
	 \kern-0.2em & a & b & c  \\
		\begin{block}{c[ccc]}
  		a \kern \rowsp & \; 1 & \underline{1} & \underline{1} \; \topstrut \\
  		b \kern \rowsp & \;  & 1 & 1 \; \\
  		c \kern \rowsp & \;  &  & 1 \; \botstrut \\
		\end{block}
	\end{blockarray}
\; \to
\begin{blockarray}{cc}{}
	\kern-0.2em & b \\
		\begin{block}{c[c]}
  		  & \; 1 \; \topstrut \\
  		 & \;  1 \; \\
  		 & \; \; \botstrut \\
		\end{block}
	\end{blockarray}
	\; \;
\begin{blockarray}{cccc}{}
	\kern-0.2em & a & b & c  \\
		\begin{block}{c[ccc]}
  		  & \; 1 &    & \underline{1} \; \topstrut \\
  		  & \;    & 1 &  1 \; \\
  		 & \;  & &  1 \; \botstrut \\
		\end{block}
	\end{blockarray}
\; \to
	\begin{blockarray}{cc}{}
	\kern-0.2em & c \\
		\begin{block}{c[c]}
  		  & \; 1 \; \topstrut \\
  		 & \; 1 \; \\
  		 & \; 1 \; \botstrut \\
		\end{block}
	\end{blockarray}
	\; \; 
	\begin{blockarray}{cccc}{}
	\kern-0.2em & a & b &  c \\
		\begin{block}{c[ccc]}
  		  & \; 1   &    &    \; \topstrut \\
  		 & \;   &  1  &   \; \\
  		 & \;   &   &  1   \; \botstrut \\
		\end{block}
	\end{blockarray}
\; \xrightarrow{P} 
\begin{blockarray}{cc}{}
	\kern-0.2em & a \\
		\begin{block}{c[c]}
  		  & \; 1 \; \topstrut \\
  		 & \; 1 \; \\
  		 & \; 1 \; \botstrut \\
		\end{block}
	\end{blockarray}
\; \;
\begin{blockarray}{cccc}{}
	\kern-0.2em & b & c & a  \\
		\begin{block}{c[ccc]}
  		  & \; 1 &  &  \; \topstrut \\
  		 & \;  & 1 &  \; \\
  		 & \; &  & 1 \; \botstrut \\
		\end{block}
	\end{blockarray}
\xrightarrow{d_V} 
\begin{blockarray}{cccc}
\kern-0.2em & b & c & a  \\
	\begin{block}{c[ccc]}
		& \;  1 & &  1 \; \topstrut \\
		& \;  & 1 &  1\; \\
		& \; &  & 1 \; \botstrut \\
	\end{block}
\end{blockarray}
$$
\noindent 
Note that the equivalent permutation using \emph{vineyards} requires $4$ column operations on both $R_1$ and $V_1$, respectively, whereas a single move operation accomplishes using only $2$ column operations per matrix. The pseudocode for \emph{MoveRight} is given in Algorithm~\ref{alg:mr} and for \emph{MoveLeft} in Algorithm~\ref{alg:ml}. 

\begin{algorithm}[!htb]
	\caption{Move Right Algorithm}\label{alg:mr}
	\setstretch{1.15}
    \begin{algorithmic}[1]
        \Function{RestoreRight}{$R$, $V$, $\mathbb{I} = \{I_1, I_2, \dots, I_s \}$, donors = $\mathrm{true}$}
            \State $(\, d_{low}, \, d_R, \, d_V \,) \gets ( \, \mathrm{low}_R(I_1), \, \mathrm{col}_R(I_1), \, \mathrm{col}_V(I_1) \, )$
        	\For{$k$ \textbf{in} $I_2, \dots, I_s$} 
        		\State $( \, d_{low}', \, d_R', \, d_V' \, ) \gets ( \, \mathrm{low}_R(k), \, \mathrm{col}_R(k), \, \mathrm{col}_V(k) \, )$
        		\State $( \, \mathrm{col}_R(k), \, \mathrm{col}_V(k) \, ) \mathrel{+}= (\, d_R, \, d_V \,) $
        		\If {$d_{low}' < d_{low}$}
        			\State $(\, d_{low}, \, d_R, \, d_V \,) \gets (\, d_{low}', \, d_R', \, d_V' \, )$
        		\EndIf
        	\EndFor 
        	\State \Return $(\, R, \, V, \, d_R, \, d_V \, )$ \textbf{if} $\mathrm{donors}=\mathrm{true}$ \textbf{else} $(\, R, \, V \, )$
        \EndFunction
    \end{algorithmic}
    \begin{algorithmic}[1]
        \Function{MoveRight}{$R$, $V$, $i$, $j$} 
            \State $\mathbb{I} =$ columns satisfying $V[i, i:j] \neq 0$ 
            \State $\mathbb{J} = $ columns satisfying $\mathrm{low}_R \in [i:j]$ and $\mathrm{row}_R(i) \neq 0$ 
            \State $(\, R, \, V, d_R, d_V \,) \gets $ \Call{RestoreRight}{$R$, $V$, $\mathbb{I}$} 
            \State $(\, R, \, V \,) \gets $ \Call{RestoreRight}{$R$, $V$, $\mathbb{J}$, $\mathrm{false}$}
            \State $(\, R, \, V\, ) \gets (\, P R P^T, \, P V P^T\,)$  
            \State $(\, \mathrm{col}_R(j), \, \mathrm{col}_V(j) \, ) \gets (\, P d_R, \, P d_V \,)$
            \State \Return $(\, R, \, V\,)$
        \EndFunction
    \end{algorithmic}
\end{algorithm}

\begin{algorithm}[!htb]
	\caption{Move Left Algorithm}\label{alg:ml}
	\setstretch{1.15}
    \begin{algorithmic}[1]
        \Function{RestoreLeft}{$R$, $V$, $\mathbb{K} = \{k_1, k_2, \dots, k_s \}$}
            \State $(l, r) \gets $ indices $l, r \in \mathbb{K}$ satisfying $l < r$, $\mathrm{low}_R(l) = \mathrm{low}_R(r)$ maximal 
            \While{$\mathrm{low}_R(l) \neq 0$ \textbf{and} $\mathrm{low}_R(r) \neq 0$}
            	\State $(\mathrm{col}_R(r), \mathrm{col}_V(r)) \mathrel{+}= (\mathrm{col}_R(l), \mathrm{col}_V(l))$
            	\State $\mathbb{K} \gets \mathbb{K} \setminus l$
            	\State $(l, r) \gets $ indices $l, r \in \mathbb{K}$ satisfying $l < r$, $\mathrm{low}_R(l) = \mathrm{low}_R(r)$ maximal
            \EndWhile
        	\State \Return $(\, R, \, V \, )$
        \EndFunction
    \end{algorithmic}
    
    \begin{algorithmic}[1]
        \Function{MoveLeft}{$R$, $V$, $i$, $j$} 
        	\State $\mathbb{I} \gets \emptyset$
            \While{$V(k, i) \neq 0 $\textbf{ for }$ k = \mathrm{low}_V(i)$\textbf{ where }$j \leq k < i$}
            	\State $(\mathrm{col}_R(i), \mathrm{col}_V(i)) \mathrel{+}= (\mathrm{col}_R(k), \mathrm{col}_V(k))$
            	\State $\mathbb{I} \gets \mathbb{I} \cup k + 1 $
            \EndWhile
            \State $(R, V) \gets (P R P^T, P V P^T)$
            \State $\mathbb{J} = $ columns satisfying $\mathrm{low}_R \in [i:j]$ and $\mathrm{row}_R(i) \neq 0$ 

            \State $(R, V) \gets$ \Call{RestoreLeft}{$R, V, \mathbb{I}$}
            \State $(R, V) \gets$ \Call{RestoreLeft}{$R, V, \mathbb{J}$}
			\State \Return $(\, R, \, V\,)$
        \EndFunction
    \end{algorithmic}
\end{algorithm}

Regarding the complexity of move operations, which clearly depend on the sparsity of $R$ and $V$, we recall the proposition shown in~\cite{busaryev2010tracking}: 
\begin{proposition}[Busaryev et al.~\cite{busaryev2010tracking}]\label{prop:move_st}
	Given a filtration with $n$ simplices of dimensions $p-1$, $p$, and $p+1$,    let $R = DV$ denote its associated decomposition. Then, the operation $\mathrm{MoveRight}(i,j)$ constructs a valid decomposition $R' = D'V'$ in $O((\lvert \mathbb{I} \rvert + \lvert \mathbb{J} \rvert)n)$ time, where $\mathbb{I}, \mathbb{J}$ are given by:
	$$ \lvert \mathbb{I}\rvert = \sum\limits_{l=i+1}^j \mathds{1}\left(\, v_l(i) \neq 0 \, \right), \quad \mathbb{J} = \sum\limits_{l=1}^m \mathds{1}\left(\, \mathrm{low}_R(l) \in [i,j] \text{ and } r_l(i) \neq 0 \, \right)$$
	Moreover, the quantity $\lvert \mathbb{I} \rvert + \lvert \mathbb{J} \rvert$ satisfies $\lvert \mathbb{I} \rvert + \lvert \mathbb{J} \rvert \leq 2(j - i)$. 
\end{proposition}
\noindent Though similar to \emph{vineyards}, move operations confer additional advantages:
 \begin{enumerate}[labelsep=5pt, topsep=1pt,itemsep=-0.25ex,parsep=1.2ex]
 	\item[M1:] Querying the non-zero status of entries in $R$ or $V$ occurs once per move.
 	\item[M2:] $R = D V$ is \underline{not} guaranteed to be valid during the movement of $\sigma_i $ to $ \sigma_j$.
 	\item[M3:] At most $O(m)$ moves are needed to reindex $f \mapsto f'$ 
 \end{enumerate} 
First, consider property M1. Prior to applying any permutation $P$ to the decomposition, it is necessary to remove non-zero entries in $V$ which render $P^TVP$ non-upper triangular, to maintain invariant I2. 
Using \emph{vineyards}, one must consistently perform $\lvert i - j \rvert - 1$ non-zero status queries interleaved between repairing column operations. A move operation groups these status queries into a single pass prior to performing any modifying operations. 

Property M2 implies that the decomposition is not fully maintained during the execution of \textit{RestoreRight} and \textit{RestoreLeft} below, which starkly contrasts the \emph{vineyards} algorithm. 
In this way, we interpret move operations as making a tradeoff in granularity: whereas a sequence of adjacent transpositions $(i, i{+}1), (i{+}1, i{+}2), \dots, (j{-}1, j)$ generates $\lvert i - j \rvert $ valid decompositions in \emph{vineyards}, an equivalent move operation $\mathrm{Move}(i,j)$ generates only one.
Indeed, Property M3 directly follows from this fact, as one may simply move each simplex $\sigma \in K$ into its new order $f'(\sigma)$ via insertion sort. Note that the number of valid decompositions produced by \emph{vineyards} is bounded above by $O(m^2)$, as each pair of simplices $\sigma_i, \sigma_j \in K$ may switch its relative ordering at most once during the interpolation from $f$ to $f'$.

As shown by example 2.3, \emph{moves} can be cheaper than \emph{vineyards} in terms of column operations. However, it is not clear that this is always the case upon inspection of Algorithm~\ref{alg:mr}, as the usage of a donor column seemingly implies that many $O(m)$ copy operations need to be performed. 
It turns out that we may handle all such operations except the first in $O(1)$ time, which we formalize below.
\begin{proposition}\label{prop:factor2}
	Let $(K,f)$ denote a filtration of size $\lvert K \rvert = m$ with decomposition $R = D V$ and let $T$ denote the number of column operations needed by \emph{vineyards} to perform the sequence of transpositions:
	$$ R = R_1 \overset{s_1}{\to} R_2 \overset{s_2}{\to} \dots \overset{s_k}{\to} R_{k-1} = R' $$
	where $s_i$ denotes the transposition $(i,i+1)$, $i < j$, and $k = \lvert i - j\rvert$. Moreover, let $M$ denote the number of column operations to perform the same update $R \mapsto R'$ with $\mathrm{Move}(i,j)$. Then the inequality $M \leq T$ holds.
\end{proposition}
\begin{proof} \normalsize
First, consider executing the \emph{vineyards} algorithm with a given pair $(i, j)$. As there are at most $2$ column operations, any contiguous sequence of transpositions $(i,i+1), (i+1,i+2), \ldots (j-1,j)$ induces at most $2(\lvert i - j \rvert)$ column operations in both $R$ and $V$, giving a total of $4(\lvert i - j \rvert)$ column operations.

Now consider MoveRight($i$,$j$), outlined in Algorithm~\ref{alg:mr}. Here, the dominant cost again are the column operations (line 5). 
Though we need an extra $O(m)$ storage allocation for the donor columns $d_\ast$ prior to the movement, notice that assignment to and from $d_\ast$ (lines (4), and (7) in \textit{RestoreLeft} and \textit{MoveRight}, respectively) requires just $O(1)$ time via a pointer swapping argument. That is, when $d_\mathrm{low}' < d_\mathrm{low}$, instead of copying $\mathrm{col}_\ast(k)$ to $d_\ast'$---which takes $O(m)$ time---we instead swap their column pointers in $O(1)$   prior to column operations. After the movement, $d_\ast$ contains the newly modified column and $\mathrm{col}_\ast(k)$ contains the unmodified donor $d_\ast'$, so the final donor swap also requires $O(1)$ time. 
Since at most one $O(m)$ column operation is required for each index in $[i, j]$, moving a column from $i$ to $j$ where $i < j$ requires at most $2(\lvert i - j \rvert)$ column operations for both $R$ and $V$. The claimed inequality follows. 
\end{proof}
As a final remark, we note that the combination of \emph{MoveRight} and \emph{MoveLeft} enable efficient simplex additions or deletions to the underlying complex. In particular, given $K$ and a decomposition $R = DV$, obtaining a valid decomposition $R' = D' V'$ of $K' = K \cup \{\sigma\}$ can be achieved by appending its requisite elementary chains to $D$ and $V$, reducing them, and then executing \emph{MoveLeft}$(m+1, i)$ with $i = f'(\sigma)$. Dually, deleting a simplex $\sigma_i$ may be achieved via \emph{MoveRight} by moving $i$-th to the end of the decomposition and dropping the corresponding columns. 


\section{Our contribution: Move Schedules}\label{sec:move_schedules}
We begin with a brief overview of the pipeline to compute the persistence diagrams of a discrete $1$-parameter family $(f_1, f_2, \dots, f_n)$ of filtrations. 
Without loss of generality, assume each filtration $f_i : K \rightarrow [m]$ is a simplexwise filtration of a fixed simplicial complex $K$ with $\lvert K \rvert =m$, and let $(f, f')$ denote any pair of such filtrations.  
Our strategy to efficiently obtain a valid $RV$ decomposition of the filtration $(K,f')$ from a given decomposition of $(K, f)$ is to decompose a fixed bijection $\rho : [m] \to [m]$ satisfying $f' = \rho \circ f$ into a \emph{schedule} of updates: 
\begin{definition}[Schedule]\label{def:schedule}
Given a pair of filtrations $(K, f), (K,f')$ 
and $R = DV$ the initial decomposition of $(K, f)$, a \emph{schedule}  $\mathcal{S} = (s_1, s_2, \dots, s_d )$ is a sequence of permutations  satisfying: 
	\begin{equation}\label{eq:rv_seq_schedule}
   		R = D_0 V_0 \overset{s_1}{\to} D_1 V_{1} \overset{s_2}{\to} \dots \overset{s_d}{\to} D_d V_{d} = R'
   	\end{equation}
   	where, for each $i \in [d]$, $R_i = D_i V_i$ is a valid decomposition respecting invariants~\ref{inv:decomposition}, and $R'$ is a valid decomposition for $(K,f')$.
\end{definition}
\noindent
To produce this sequence of permutations $\mathcal{S} = (s_1, \dots, s_d)$ from $\rho$, our approach is as follows:
define $q = (\rho(1), \rho(2), \ldots, \rho(m) )$. 
We compute a longest increasing subsequence $\mathrm{LIS}(q)$, and then use this subsequence to recover a longest common subsequence (LCS) between $f$ and $f'$, which we denote later with $\mathrm{LCS}(f, f')$. 
We pass $q$, $\mathrm{LIS}(q)$, and a ``greedy'' heuristic for picking moves $h$ to our scheduling algorithm (to be defined later), which returns as output an ordered set of move permutations  $\mathcal{S}$ of minimum size satisfying~\eqref{eq:rv_seq_schedule}---a \emph{schedule} for the pair $(f, f')$.
These sequence of steps is outlined in Algorithm~\ref{alg:schedule} below.

\begin{algorithm}[h]
	\caption{Scheduling algorithm}\label{alg:schedule}
	\setstretch{1.05}
    \begin{algorithmic}[1]
    	\Require Filtration pair $\mathcal{F} = \{f_0, f_1\}$ and valid $R = DV$ for $(K,f_0)$
    	\Ensure Valid $R = D V$ is computed for $(K, f_1)$
    	\Procedure{MoveSchedule}{$\mathcal{F} = ( f_0, f_1 )$, $R$, $V$} 
    		\State Fix $\rho \in S_m$ such that $f_1 = \rho \circ f_0$ \Comment{$O(m)$} 
    		\State $q \gets  (\rho(1), \rho(2), \ldots, \rho(m) )$ \Comment{$O(m)$}
    		\State $\mathrm{lis}_{q} \gets \mathrm{LIS}(q)$ \Comment{$O(m \log \log m)$}
			\State $h \gets \texttt{greedy}$ \Comment{Heuristic for picking moves}
    		\State $\mathcal{S} \gets \mathrm{Schedule}(q, \mathrm{lis}_{q}, h)$ \Comment{$O(d^2\log m)$, see Section~\ref{sec:schedule_cost}}
	    	\For{$(i,j)$\textbf{ in }$\mathcal{S}$} \Comment{$\lvert \mathcal{S} \rvert = d$}
	    		\State $(R, V) \gets$ \textbf{if} $i < j$ \Call{MoveRight}{i, j} \textbf{else} \Call{MoveLeft}{i, j}
	    	\EndFor
    	\EndProcedure
	\end{algorithmic}
\end{algorithm}
Though using the LCS between a pair of permutations induced by simplexwise filtrations is a relatively intuitive way of producing a schedule of move-type decomposition updates, it is not immediately clear whether such an algorithm has any computational benefits compared to \emph{vineyards}. 
Indeed, since \emph{moves} is a generalization of \emph{vineyards}, several important questions arise when considering practical aspects of how to implement Algorithm~\ref{alg:schedule}, such as e.g. how to pick a ``good'' heuristic $h$, how expensive can the $\text{Schedule}$ algorithm be, or how  $\lvert \mathcal{S}\rvert$ scales with respect to the size of the complex $K$. We address these issues in the following sections. 


\subsection{Continuous setting}\label{sec:continuous_setting}
In the \emph{vineyards} setting, a given homotopy  $F : K \times [0,1] \to \mathbb{R}$ continuously interpolating between $(K, f)$ and $(K, f')$ is discretized into a set of critical events that alter the filtration order. 
As $F$ determines the number of distinct filtrations encountered during the deformation from $f$ to $f'$, a natural question is whether such an interpolation can be modified so as to minimize $\lvert \mathcal{S}\rvert$---the number of times the decomposition is restored to a valid state.
Towards explaining the phenomenon exhibited in Figure~\ref{fig:vineyards}, we begin by analyzing a class of interpolation schemes to establish an upper bound on this quantity.


Let $F : K \times [0,1] \to \mathbb{R}$ be a homotopy of $x$-monotone curves\footnote{This term has been used in reference to parameterized curves whose behavior with respect to a certain restricted set of geometric predicates is invariant, see~\cite{boissonnat2000efficient}. } between the  filtrations 
$f,f' : K \rightarrow [m]$ whose function $t\mapsto F(\sigma, t)$ is continuous and satisfies $f(\sigma) = F(\sigma, 0)$ and $f'(\sigma) = F(\sigma, 1)$ for every  $\sigma \in K$.
Note that this family includes the straight-line homotopy $F(\sigma, t) = (1 - t) f(\sigma) + t f'(\sigma)$,  studied in the original \emph{vineyards} paper~\cite{cohen2006vines}. 
If we assume that each pair of curves $\big(t,  F(\cdot, t)\big) \subset [0,1]\times \mathbb{R}$  intersect in at most one point---at which they cross---the continuity and genericity assumptions on $F$ imply that for $ \sigma,\mu \in K$ distinct,  the curves   $t \mapsto F(\sigma ,t)$ and $t \mapsto F(\mu,t)$ intersect if and only if $f(\sigma) > f(\mu)$ and $f'(\sigma)  < f'(\mu)$, or $f(\sigma) < f(\mu)$ and $f'(\sigma)  > f'(\mu)$. 
 In other words, the number of crossings in $F$ is 
 exactly the  \emph{Kendall}-$\tau$ distance~\cite{diaconis1977spearman} between $f$ and $f'$: 
\begin{equation}\label{eq:kendall_dist}
	K_\tau(f, f') = 
 \frac{1}{2} \left\lvert \, \left\{(\sigma, \mu) \mid \mathsf{sign}\big(f(\sigma) - f(\mu) \big) \neq \mathsf{sign}\big(f'(\sigma) - f'(\mu)\big) \right\} \, \right\rvert
\end{equation}
After slightly perturbing $F$ if necessary,  we can further assume that its crossings occur at  $k = K_\tau(f,f')$ distinct time points $0< t_1 < \cdots < t_k < 1$. 
Let $t_0 = 0$, $t_{k+1}= 1$ and fix 
$a_i \in (t_{i}, t_{i + 1})$ for $i=0, \ldots, k$. 
Then, the order in $K$ induced by $\sigma \mapsto F(\sigma, a_i)$
defines a filtration $f_i : K \rightarrow [m]$
so  that $f_0 = f$, $f_k = f'$ and 
$\mathcal{F} = (f_0, f_1,\ldots, f_k)$ is the ordered sequence of all distinct filtrations in the interpolation from $f$ to $f'$ via $F$.

The continuity of the curves $t\mapsto F(\cdot, t)$ and the fact that $t_i$ is the sole crossing time  in the interval $(t_{i-1}, t_{i+1})$, imply that the
permutation $\rho_i $   
transforming $f_{i-1}$ into $f_i$, i.e. so that  $f_i = \rho_i \circ f_{i-1}$, must be (in cycle notation) of the form 
$\rho_i = (\ell_i \;\; \ell_i +1 )$ for $1 \leq \ell_i < m$. In other words, $\rho_i$ is an adjacent transposition for each $i =1,\ldots, k$.
Observe the size of the ordered sequence of adjacent transpositions  $S_F = (\rho_1, \rho_2, \ldots, \rho_k)$ defined from the homotopy $F$ above is exactly $K_\tau(f,f')$.
On the positive side, the reduction of schedule planning to crossing detection implies the former can be solved optimally in output-sensitive $O(m \log m + k)$ time by several algorithms~\cite{boissonnat2000efficient}, where $k$ is the output-sensitive term and $m$ is the number of simplices in the filtration(s).
On  the negative side, $k = K_\tau(f,f')$ scales in size to $\sim O(m^2)$ in the worst case, achieved when $f' = - f$. As mentioned in~\ref{sec:vineyards}, this quadratic scaling induces a number of issues in the practical implementations of the \emph{vineyards} algorithm

\begin{remark}
	The grayscale image data example from section~\ref{sec:motivation} exhibits this quadratic scaling.
Indeed, the Freudenthal triangulation of the $9\times 9$ grid contains $(81, 208, 128)$ simplices of dimensions $(0, 1, 2)$, respectively. 
Therefore, $m = 417$ and $\lvert S_F \rvert \leq \frac{1}{2}m(m-1) = 86,736$.
As the homotopy given by the video is varied,  $\approx 70,\!000$ transpositions are generated, approaching the worst case upper bound due to the fact that $f'$  is nearly the reverse of $f$.  
\end{remark}

If our goal is to decrease  $\lvert S_F \rvert$, one option is to \emph{coarsen} $S_F$ to a new schedule $\widetilde{S}_F$ by e.g. collapsing contiguous sequences of adjacent transpositions to moves, via the map $(i, i+1)(i+1, i+2)\cdots(j-1, j) \mapsto (j, i+1, \cdots, j-1, i)$ (if $i < j$).
Clearly $\lvert \widetilde{S}_F \rvert \leq \lvert S_F \rvert$ and the associated coarsened $\widetilde{S}_F$ requires just $O(m)$ time to compute. However, the coarsening depends entirely on the initial choice of $F$ and the quadratic upper bound remains---it is always possible that there are no contiguous subsequences to collapse. 
This suggests one must either abandon the continuous  setting or make stronger assumptions on $F$ to have any hope of keeping $\lvert S_F \rvert \sim O(m)$ in size.

\subsection{Discrete setting}\label{sec:moves_setting}


Contrasting the continuous-time setting, if we discard the use of a homotopy interpolation and allow move operations in any order, we obtain a trivial upper bound of $O(m)$ on the schedule size:  simply move each simplex in $K$ from  its position in the filtration given by $f$ to the position given by $f'$---which we call the \emph{naive strategy}. In particular, in losing the interpolation interpretation, it is no longer clear the $O(m)$ bound is tight. Indeed, the ``intermediate'' filtrations need no longer even respect the face poset of the underlying complex $K$. 
In this section, we investigate these issues from a combinatorial perspective. 

Let $S_m$ denote the symmetric group. Given two fixed permutations $p, q \in S_m$ and a set of allowable permutations $\Sigma \subseteq S_m$,  a common  problem is to find a sequence of permutations $s_1, s_2, \dots, s_d \in \Sigma$ whose composition satisfies:
	\begin{equation}\label{eq:sorting}
		s_d \circ \cdots \circ s_2 \circ s_1 \circ p = q
	\end{equation}
Common variations of this problem include finding such a sequence of minimal length ($d$) and bounding the length $d$ as a function of $m$. In the latter case, the largest  lower bound on $d$ is referred to as the \emph{distance} between $p$ and $q$ with respect to $\Sigma$. 
A sequence $S = (s_1, s_2, \dots, s_d)$ of operations $s \in \Sigma \subseteq S_m$ mapping $p \mapsto q$ is sometimes called a \emph{sorting of $p$}. 
When $p, q$ are interpreted as strings, these operations $s \in \Sigma$ are called \emph{edit operations}. The minimal number of edit operations $d_\Sigma(p, q)$ needed to sort $p \mapsto q$ with respect to $\Sigma$ is referred to as the \emph{edit distance}~\cite{bergroth2000survey} between $p$ and $q$.
We denote the space of sequences transforming $p \mapsto q$ using $d$ permutations in $\Sigma \subseteq S_m$ with $\Phi_\Sigma(p,q,d)$.
Note the choice of $\Sigma$ defines the type of distance being measured---otherwise if $\Sigma = S_m$, then $d_\Sigma(p, q) = 1$ trivially for any $p\neq  q \in S_m$.

Perhaps surprisingly, small changes to the set of allowable edit operations $\Sigma$ dramatically affect both the size of $d_\Sigma(p,q)$ and the difficulty of obtaining a minimal sorting. 
For example, while sorting by transpositions and reversals is NP-hard and sorting by prefix  transpositions is unknown, there are polynomial time algorithms for sorting by block interchanges, exchanges, and prefix exchanges~\cite{labarre2013lower}. Sorting by adjacent transpositions can be achieved in many ways: any sorting algorithm that exchanges two adjacent elements during its execution (e.g. bubble sort, insertion sort) yields a sorting of size $K_\tau(p, q)$.  

Here we consider sorting by moves. Using permutations, a \emph{move operation} $m_{ij}$ that moves $i$ to $j$ in $[m]$, for $i < j$, corresponds to the circular rotation: 
\begin{equation}\label{eq:move_perm}
m_{ij} = \setlength\arraycolsep{1.8pt}
\scalefont{0.90}{
\setcounter{MaxMatrixCols}{20}
\left (
\begin{array}{*3c | *5c | *{3}c } 
\cline{4-8} \rule{0pt}{1ex}
1 & \cdots & i -1 & i & i + 1 & \cdots & j - 1 & j & j +1 &  \cdots & m \\
\cline{4-8}
\cline{4-8}\rule{0pt}{1ex}
1 & \cdots & i -1 & i + 1 & \cdots & j - 1 & j & i & j +1 &  \cdots & m \\
\cline{4-8}
\end{array}
\right ) 
}
\end{equation}
In cycle notation, this corresponds to the cyclic permutation $m_{ij} = (\enspace i \enspace j \enspace j\text{-}1 \enspace \dots \enspace i\text{+}2 \enspace i\text{+}1 \enspace )$.
Observe that a move operation can be interpreted as a paired delete-and-insert operation, i.e. $m_{ij} = (\mathrm{ins}_j \circ \mathrm{del}_i)$, where $\mathrm{del}_i$ denotes the operation that deletes the character at position $i$  and $\mathrm{ins}_j$ the operation that inserts the same character at position $j$. 
Thus, sorting by move operations can be interpreted as finding a minimal sequence of edits where the only operations allowed are (paired) insertions and deletions---this is exactly the well known \emph{Longest Common Subsequence} (LCS) distance. Between strings $p, q$ of sizes $m$ and $n$, the LCS distance is given by~\cite{bergroth2000survey}: 
\begin{equation}
	\mathrm{d}_{\mathrm{lcs}}(p,q) = m + n - 2\lvert \mathrm{LCS}(p, q)\rvert
\end{equation}
%
With this insight in mind, we obtain the following bound on the minimum size of a sorting (i.e. schedule) using moves and the complexity of computing it. 
\begin{proposition}[Schedule Size]\label{prop:schedule_size}
Let $(K, f), (K, f')$ denote two filtrations of size $\lvert K \rvert = m$. Then, the smallest move schedule $S^*$ re-indexing $f \mapsto f'$ has size: 
$$ \lvert S^\ast \rvert = d = m - \lvert \mathrm{LCS}(f, f') \rvert $$ 
where we use $\mathrm{LCS}(f, f')$ to denote the LCS of the permutations of $K$ induced by $f$ and $f'$.
\end{proposition}
\begin{proof} \normalsize
Recall our definition of  edit distance given above, 
depending on the choice  $\Sigma \subseteq S_m$ of allowable edit operations, and that in order for any edit distance to be symmetric, if $s \in \Sigma$ then $s^{-1} \in \Sigma$. This implies that $d_\Sigma(p,q) = d_\Sigma(p^{-1}, q)$ for any choice of  $p,q \in S_m$. 
Moreover, edit distances are \emph{left-invariant}, i.e.
\[
d_\Sigma(p,q) = d_\Sigma(r \circ p, r \circ q) \quad \text{ for all } p,q,r \in S_m
\]
Conceptually, left-invariance implies that the edit distance between any pair of permutations $p,q$ is invariant under an arbitrary relabeling of $p,q$--as long as the relabeling is consistent. Thus, the following identity always holds: 
$$ d_\Sigma(p,q) = d_\Sigma(\iota, p^{-1} \circ q) = d_\Sigma(q^{-1} \circ p, \iota) $$
where $\iota = [m]$, the identity permutation. Suppose we are given two permutations $p, q \in S_n$ and we seek to compute $\mathrm{LCS}(p, q)$. Consider the permutation $p' = q^{-1} \circ p$. Since the LCS distance is a valid edit distance, if $\lvert \mathrm{LCS}(p, q) \rvert = k$, then $\lvert \mathrm{LCS}(p', \iota) \rvert = k$ as well. Notice that $\iota$ is strictly increasing, and that any common subsequence $p'$ has with $\iota$ must also be strictly increasing. The optimality of $d$ follows from the optimality of the well-studied LCS problem~\cite{kumar1987linear}. 
\end{proof}
\noindent
For any pair of general string inputs of sizes $n$ and $m$, respectively, the LCS between them is computable in $O(mn)$ with dynamic programming, and there is substantial evidence that the complexity cannot be much lower than this~\cite{abboud2015tight}. 
In our setting, however, we interpret the simplexwise filtrations $(K, f), (K, f')$ as \emph{permutations} of the same underlying complex $K$---in this setting, $d_{\mathrm{LCS}}$ reduces further to the \emph{permutation edit distance} problem. 
This special type of edit distance has additional structure to it, which we demonstrate below.  
\begin{corollary}\label{cor:lis_schedule_complexity}
Let $(K, f), (K, f')$ denote two filtrations of size $\lvert K \rvert = m$, and let $S^*$ denote schedule of minimal size re-indexing $f \mapsto f'$. Then $\lvert S^\ast \rvert = d$ can be determined in $O(m \log \log m)$ time. 
\end{corollary}
\begin{proof} \normalsize
	By the same reduction from Proposition~\ref{prop:schedule_size}, the problem of computing $\mathrm{LCS}(f, f')$ reduces to the problem of computing the \emph{longest increasing subsequence} (LIS) of a particular permutation $p' \in S_m$, which can be done in $O( m \log \log m)$ time using van Emde Boas trees~\cite{bespamyatnikh2000enumerating}
\end{proof}
\noindent 
Reduced complexity is not the only immediate benefit from Proposition~\ref{prop:schedule_size}; by the same reduction to the LIS problem, we obtain the worst-case bounds on $\mathcal{S}$ in expectation. 
\begin{corollary}\label{cor:expectation}
If $(K, f), (K,f')$ are random filtrations of a common complex $K$ of size $m$, then the expected size of a longest common subsequence $\mathrm{LCS}(f, f')$ between $f,f'$ is no larger than $m - \sqrt{m}$, with probability $1$ as $m \to \infty$. 
\end{corollary}
\begin{proof} \normalsize
	The proof of this result reduces to showing the average length of the LIS for random permutations. Let $L(p) \in [1,m]$ denote the maximal length of a increasing subsequence of $p \in S_m$. 
	The essential quantity to show the expected length of $L(p)$ over all permutations: 
	$$ \mathbb{E} \, L(p) = \ell_m = \frac{1}{m!} \sum\limits_{p \in S_m} L(p)$$
	A large body of work dates back at least 50 years has focused on estimating this quantity, which is sometimes called the \emph{Ulam-Hammersley} problem. Seminal work by Baik et al.~\cite{baik1999distribution} established that as $m \to \infty$:
	$$ \displaystyle \ell_m = 2 \sqrt{m} + c m^{1/6} + o(m^{1/6}) $$
where $c = -1.77108...$. Moreover, letting $m \to \infty$, we have: 
$$ \frac{\ell_m}{\sqrt{m}} \to 2\rlap{\quad \text{as } $m \to \infty$} $$	 
Thus, if $p \in S_m$ denotes a uniformly random permutation in $S_m$, then $L(p)/\sqrt{m} \to 2$ in probability as $m \to \infty$. Using the reduction from above to show that $\mathrm{LCS}(p,q) \Leftrightarrow \mathrm{LIS}(p')$, the claimed bound follows.
\end{proof}
\begin{remark}
\noindent Note the quantity from Corollary~\ref{cor:expectation} captures the size of $S^\ast$ between pairs of uniformly sampled permutations, as opposed to uniformly sampled filtrations, which have more structure due to the face poset. 
However, Boissonnat~\cite{boissonnat2018efficient} prove the number of distinct filtrations built from a $k$-dimensional simplicial complex $K$ with $m$ simplices and $t$ distinct filtration values is \emph{at least} $\floor{\frac{t+1}{k+1}}^m$. 
Since this bound grows similarly to $m!$ when $t \sim O(m)$ and $k << m$ fixed, $d \approx n - \sqrt{n}$ is not too pessimistic a bound between random filtrations.     
\end{remark}
\noindent 
In practice, when one has a time-varying filtration and the sampling points are relatively close [in time], the LCS between adjacent filtrations is expected to be much larger, shrinking $d$ substantially. 
For example, for the complex from Section~\ref{sec:motivation} with $m=417$ simplices, the average size of the LCS across the 10 evenly spaced filtrations was $343$, implying $d \approx 70$ permutations needed on average to update the decomposition between adjacent time points.
 
We conclude this section with the main theorem of this effort: an output-sensitive bound on the computation of persistence dynamically.
  \begin{theorem}\label{prop:sparsity_result}
   Given a pair of filtrations $(K,f), (K,f')$, a decomposition $R = DV$ of $K$, and a sequence $\mathcal{S} = \left( s_1, s_2, \dots, s_d \right)$ of cyclic `move' permutations $s_k = (i_k, j_k)$ satisfying $i_k < j_k$ for all $k \in [d]$, computing the updates:
   \begin{equation}\label{eq:rv_seq_theorem}
   		R = D_0 V_0 \overset{s_1}{\to} D_1 V_{1} \overset{s_2}{\to} \dots \overset{s_d}{\to} D_d V_{d} = R'
   	\end{equation}
   	where $R' = D_d V_d$ denotes a valid decomposition of $(K, f')$ requires $O(\nu)$ column operations, where $\nu$ depends on the sparsity of the intermediate entries $V_1, V_2, \dots, V_d$ and $R_1, R_2, \dots, R_d$:	  
   	$$  \quad \nu = \sum\limits_{k = 1}^d \, (\lvert \mathbb{I}_{k} \rvert + \lvert \mathbb{J}_{k} \rvert), \quad \text{ where }  \lvert \mathbb{I}_{k} \rvert, \lvert \mathbb{J}_{k} \rvert \text{are given in Proposition~\ref{prop:move_st} } $$ 
   	Moreover, the size of a minimal such $\mathcal{S}$ can be determined in $O(m \log \log m)$ time and $O(m)$ space. 
\end{theorem}
\begin{proof} \normalsize
	Proposition~\ref{prop:schedule_size} yields the necessary conditions for constructing $\mathcal{S}$  with optimal size $d$ in $O(m \log \log m)$ time and $O(m)$. The definition of $\nu$ follows directly from Algorithm~\ref{alg:mr}. 
\end{proof}

\subsection{Constructing schedules}\label{sec:schedule_construction}
While it is clear from the proof in Proposition~\ref{prop:schedule_size} that one may compute the LCS between two permutations $p, q \in S_m$ in $O(m \log \log m)$ time, it is not immediately clear how to obtain a sorting $p \mapsto q$ from a given $\mathcal{L} = \mathrm{LCS}(p,q)$ efficiently. We outline below a simple procedure which constructs such a sorting $\mathcal{S} = (s_1, \dots, s_d)$ in $O(dm\log m)$ time and $O(m)$ space, or $O(m \log m)$ time and $O(m)$ space per update in the online setting.
 
Recall that a \emph{sorting} $\mathcal{S}$ with respect to two permutations $p, q \in S_m$ is an ordered sequence of permutations $\mathcal{S} = (s_1, s_2, \dots, s_d)$ satisfying $q = s_d \circ \dots s_1 \circ p$.
By definition, a subsequence in $\mathcal{L}$ common to both $p$ and $q$ satisfies:
\begin{equation}
	p^{-1}(\sigma) < p^{-1}(\tau) \implies q^{-1}(\sigma) < q^{-1}(\tau) \quad \forall \sigma, \tau \in \mathcal{L}
\end{equation}
 where $p^{-1}(\sigma)$ (resp. $q^{-1}(\sigma)$) denotes the position of $\sigma$ in $p$ (resp. $q$). Thus, obtaining a sorting $p \mapsto q$ of size $d = m - \lvert \mathcal{L} \rvert$ reduces to applying a sequence of moves in the complement of $\mathcal{L}$. 
 Formally, we define a permutation $s \in S_m$ as a \emph{valid} operation with respect to a fixed pair $p, q \in S_m$ if it satisfies:
 \begin{equation}\label{eq:valid_move}
 	\lvert \mathrm{LCS}(s \circ p, q) \rvert = \lvert \mathrm{LCS}(p,q) \rvert + 1
 \end{equation}
  The problem of constructing a sorting $\mathcal{S}$ of size $d$ thus reduces to choosing a sequence of $d$ valid moves, which we call a \emph{valid sorting}. 
  To do this efficiently, let $\mathcal{U}$ denote an ordered set-like data structure that supports the following operations on elements $\sigma \in M$ from the set $M = \{0, 1, \dots, m+1\}$:
  \begin{enumerate}
  	\item $\mathcal{U} \cup \sigma$---inserts $\sigma$ into $\mathcal{U}$,
  	\item $\mathcal{U} \setminus \sigma$---removes $\sigma$ from $\mathcal{U}$,
  	\item $\mathcal{U}_{\mathrm{succ}}(\sigma)$---obtain the successor of $\sigma$ in $\mathcal{U}$, if it exists, otherwise return $m+1$
  	\item $\mathcal{U}_{\mathrm{pred}}(\sigma)$---obtain the predecessor of $\sigma$ in $\mathcal{U}$, if it exists, otherwise return $0$
  \end{enumerate}
  Given $\mathcal{U}$, a valid sorting can be constructed by querying and maintaining information about the $\mathrm{LCS}$ in $\mathcal{U}$. To see this, suppose $\mathcal{U}$ contains the current $\mathrm{LCS}$ between two permutations $p$ and $q$. By definition of the LCS, we have:
 \begin{equation}\label{eq:valid_move_ineq}
 	p^{-1}(\mathcal{U}_{\mathrm{pred}}(\sigma)) < p^{-1}(\sigma) < p^{-1}(\mathcal{U}_{\mathrm{succ}}(\sigma))
 \end{equation}
  for every $\sigma \in \mathcal{U}$. Now, suppose we choose some element $\sigma \notin \mathcal{U}$ which we would like to add to the LCS. If $p^{-1}(\sigma) < p^{-1}(\mathcal{U}_{\mathrm{pred}}(\sigma))$, then we must move $\sigma$ to the right of its predecessor in $p$ such that~\eqref{eq:valid_move_ineq} holds. Similarly, if  $p^{-1}(\mathcal{U}_{\mathrm{succ}}(\sigma)) < p^{-1}(\sigma)$, then we must move $\sigma$ left of its successor in $p$. Assuming the structure $\mathcal{U}$ supports all of the above operations in $O(\log m)$ time, we easily deduce a $O(d m\log m)$ algorithm for obtaining a valid sorting.

\subsection{Minimizing schedule cost}\label{sec:schedule_cost}
The algorithm outlined in section~\ref{sec:schedule_construction} is a sufficient for generating move schedules of minimal cardinality: any schedule of moves $S$ sorting $f \mapsto f'$ above is guaranteed to have size $\lvert S \rvert = m - \lvert \mathrm{LCS}(f, f') \rvert$, and the reduction to the permutation edit distance problem ensures this size is optimal. 
However, as with the \emph{vineyards} algorithm, certain pairs of simplices cost more to exchange depending on whether they are critical pairs in the sense described in~\cite{cohen2006vines}, resulting in a large variability in the cost of randomly generated schedules. This variability is undesirable in practice: we would like to generate a schedule which not only small in size, but is also efficient in terms of its required column operations.

\subsubsection{Greedy approach}\label{sec:greedy} 
Ideally, we would like to minimize the cost of a schedule $\mathcal{S} \in \Phi_\Sigma(p,q,d)$ directly, which recall is given by the number of non-zeros at certain entries in $R$ and $V$: 
\begin{equation}\label{eq:schedule_cost}
\mathrm{cost}(\mathcal{S}) = \sum\limits_{k=1}^d \lvert \mathbb{I}_k \rvert + \lvert \mathbb{J}_k\rvert
\end{equation}
where $\lvert \mathbb{I}\rvert + \lvert \mathbb{J}\rvert$ are the quantities from Proposition~\ref{prop:move_st}. 
Globally minimizing the objective~\eqref{eq:schedule_cost} directly is difficult due to the changing sparsity of the intermediate matrices $R_k, V_k$. 
One advantage of the \emph{moves} framework is that the cost of a single move on a given $R = D V$ decomposition can be determined efficiently prior to any column operations. Thus, it is natural to consider whether one could minimize~\eqref{eq:schedule_cost} by greedily choosing the lowest cost move in each step. Unfortunately, not only does this approach does not yield a minimal cost solution, we give a counter-example in the appendix (\ref{sec:greedy_counter_ex}) demonstrating such a greedy procedure may lead to arbitrarily bad behavior. 

\subsubsection{Proxy objective}\label{sec:proxy_objective}

In light of section~\ref{sec:greedy}, we seek an alternative objective that correlates with~\eqref{eq:schedule_cost} and does not depend on the entries in the decomposition.
Given a pair of filtrations $(K, f), (K,f')$, a natural schedule $\mathcal{S} \in \Phi(f,f',d)$ of cyclic permutations $(i_1, j_1), (i_2, j_2), \dots, (i_d, j_d)$ is one minimizing the upper bound:
\begin{equation}\label{eq:upper_bound_move_cost}
\mathrm{\widetilde{cost}}(\mathcal{S}) = \sum\limits_{k=1}^d 2 \lvert i_k - j_k \rvert \geq \sum\limits_{k=1}^d ( \lvert \mathbb{I}_k \rvert + \lvert \mathbb{J}_k\rvert )
\end{equation}
Unfortunately, even obtaining an optimal schedule $\mathcal{S}^\ast \in \Phi(f,f',d)$ minimizing~\eqref{eq:upper_bound_move_cost}
does not appear tractable due to its similarity with the $k$-layer crossing minimization problem, which is NP-hard for $k$ sets of permutations when $k \geq 4$~\cite{biedl2009complexity}. For additional discussion on the relationship between these two problems, see section~\ref{sec:cross_minimization} in the appendix.

In light of the discussion above, we devise a \emph{proxy} objective function based on the Spearman distance~\cite{diaconis1977spearman} which we observed is both efficient to optimize and effective in practice. 
The \emph{Spearman footrule distance} $F(p,q)$ between two $p,q \in S_m$ is an $\ell_1$-type distance for measuring permutation disarrangement:
\begin{equation}\label{eq:spearman_dist}
F(p, q) = \sum\limits_{i =1}^m \lvert \, p(i) - q(i) \, \rvert =  \sum\limits_{i =1}^m \lvert \,i - (q^{-1} \circ p)(i) \, \rvert
\end{equation}
Like $K_\tau$, $F$ forms a metric on $S_m$ and is invariant under relabeling.
Our motivation for considering the footrule distance is motivated by the fact that $F$ recovers $\mathrm{\widetilde{cost}}(\mathcal{S})$ when $(p,q)$ differ by a cyclic permutation (i.e. $F(p, m_{ij} \circ p) = 2 \lvert i - j \rvert$) and by its usage on similar\footnote{
$F$ is often used to approximate $K_\tau$ in \emph{rank aggregation} problems due to the fact that computing the [Kemini] optimal rank aggregation is NP-hard with respect to $K_\tau$, but only polynomial time with respect to $F$~\cite{dinu2006efficient}. Note $F$ bounds $K_\tau$ via the inequalities $K_\tau(p, q) \leq F(p, q) \leq 2 K_\tau(p, q)$~\cite{diaconis1977spearman}.
} combinatorial optimization problems. 
To adapt $F$ to sortings, we decompose $F$ additively via the bound:
\begin{equation}\label{eq:sorting_additive}
\hat{F}_{\mathcal{S}}(p, \iota) = \sum\limits_{i=0}^{d-1} F(\hat{s}_i \circ p, \hat{s}_{i+1} \circ p) \geq F(p, \iota)
\end{equation}
where $\hat{s}_i =  s_{i} \circ \cdots \circ s_2 \circ s_1$ denotes the composition of the first $i$ permutations of a sorting  $S = (s_1, \dots, s_d)$ that maps $p \mapsto \iota$ and $s_0 = \iota$.
As a heuristic to minimize~\eqref{eq:sorting_additive}, we greedily select the optimal $k \in \mathcal{L}$ at each step which minimizes the Spearman distance to the identity permutation: 
\begin{equation}\label{eq:greedy_step}
	k_{\text{greedy}} = \argmin_{k \in \mathcal{D}} F(m_{ij} \circ p, \iota), \quad i = p^{-1}(k)
\end{equation}  
Note that equality between $F$ and $\hat{F}_{\mathcal{S}}$~\eqref{eq:sorting_additive} is achieved when the displacement of each $\sigma \in p$ between its initial position in $p$ to its value is non-increasing with every application of $s_i$, which in general not guaranteed using the schedule construction method in section~\ref{sec:move_schedules}. 
To build intuition for how this heuristic interacts with Algorithm~\ref{alg:sorting}, we show a purely combinatorial example below.
\\
\\  
\noindent \textbf{Example:} The permutation $p \in S_m$ to sort to the identity $p \mapsto \iota$ is given as a sequence $(4, 2, 7, 1, 8, 6, 3, 5,0)$ and a precomputed LIS $\mathcal{L} = (1, 3, 5)$, which is highlighted in red. 
\begin{figure}[htb!]\label{fig:schedule_construct}
	\center
	\includegraphics[width=0.80\textwidth]{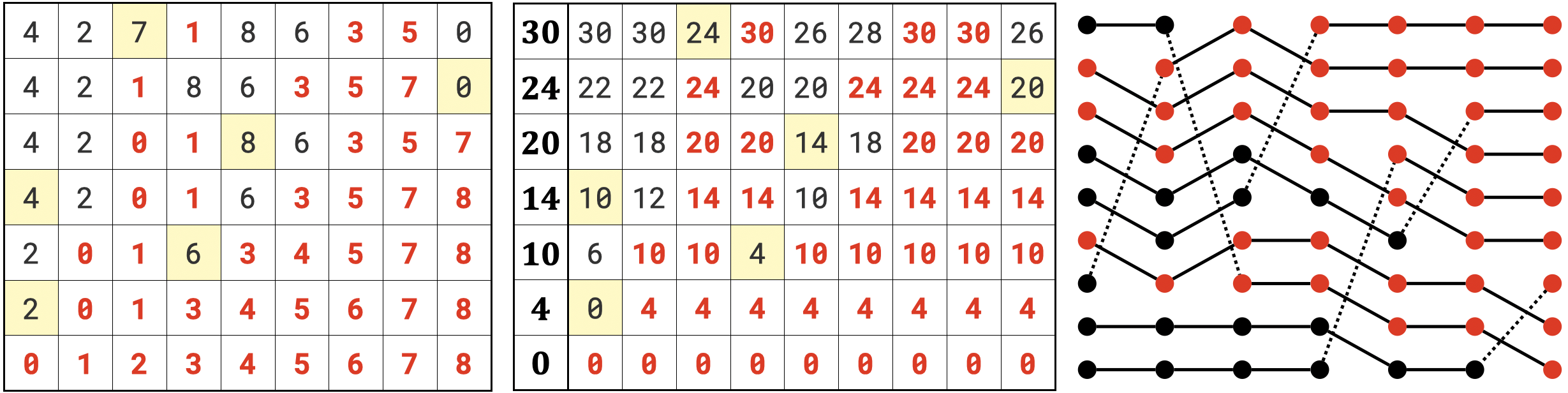} 
\end{figure}
On the left, a table records permuted sequences $\{ \hat{s}_i \circ p \}_{i=0}^d$, given on rows $i \in \{0, 1, \dots, 6\}$, for the moved elements $(7, 0, 8, 4, 6, 2)$ (highlighted in yellow). 
 In the middle, we show the Spearman distance cost~\eqref{eq:greedy_step} associated with both each permuted sequence (left column) and with each candidate permutation $m_{ij} \circ p$ (rows). Note that elements $\sigma \in \mathcal{L}$ (red) induce identity permutations that do not modify the cost, and thus the minimization is restricted to elements $\sigma \in p \setminus \mathcal{L}$ (black). 
 The right plot shows the connection to the bipartite crossing minimization problem discussed in section~\ref{sec:greedy}.

\subsubsection{Heuristic Computation}
We now introduce an efficient $O(d^2 \log m)$ algorithm for constructing a move schedule that greedily minimizes~\eqref{eq:sorting_additive}. The algorithm is purely combinatorial, and is motivated by the simplicity of updating the Spearman distance between sequences which differ by a cyclic permutation. 

First, consider an array $\mathcal{A}$ of size $m$ which provides $O(1)$ access and modification, initialized with the \emph{signed displacement} of every element in $p$ to its corresponding position in $q$. 
In the case where $q = \iota$, note $\mathcal{A}(i) = i - p(i)$, thus $F(p, \iota)$ is given by the sum of the absolute values of the elements in $\mathcal{A}$.
In other words, computing $F(m_{ij} \circ p, \iota)$ in $O(\log m)$ time corresponds to updating an array's \emph{prefix sum} under element insertions and deletions, which is easily solved by many data structures (e.g. segment trees).

\begin{figure}[htb!]
	\center
	\caption{We re-use the previous example of sorting the sequence $p = (4, 2, 7, 1, 8, 6, 3, 5,0)$ to the identity $\iota = [m]$ using a precomputed LIS $\mathcal{L} = (1, 3, 5)$. In the left table shown below, we display the same sorting as before, with elements chosen to move highlighted in yellow; on the right, a table showing the entries of $\mathcal{A}$ are recorded. }
	\includegraphics[width=0.70\textwidth]{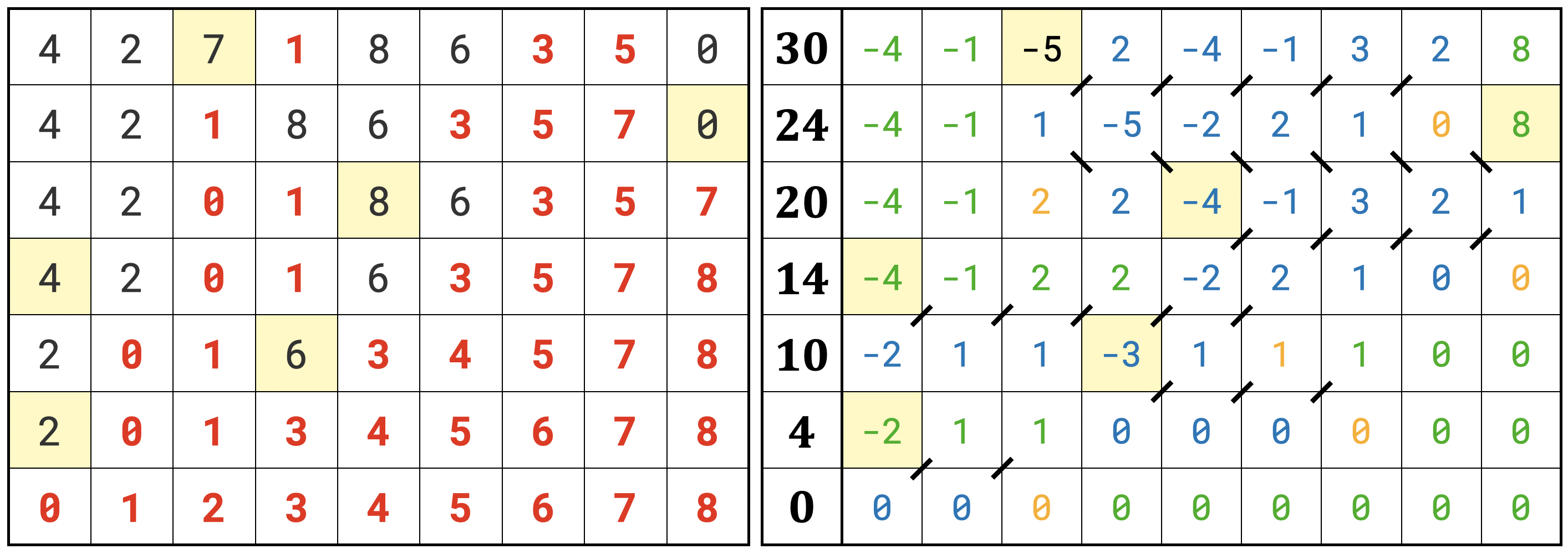}
	\caption*{The colors in the right table group the range updates: green values are unchanged between rows, blue values are modified (and shifted) by $\pm 1$, and orange values are modified arbitrarily. As before, the $i$th entry of the column on the left-side shows $F(\hat{s}_i \circ p, \iota)$. Small black lines are used to show how the entries in $\mathcal{A}$ which change by $\pm 1$ each move. }\label{fig:displacement_proof}
\end{figure}

Because the values of $\mathcal{A}$ are signed, the reduction to prefix sums is not exact: it is not immediately clear how to modify $\lvert i - j \rvert$ elements in $O(\log m)$ time. 
To address this, observe that at any point during the execution of Algorithm~\ref{alg:sorting}, a cyclic permutation changes each value of $\mathcal{A}$ in at most three different ways: 
\[
\mathcal{A}(m_{ij} \circ p) = 
\begin{cases} 
	 \mathcal{A}(k) \pm \lvert i - j \rvert & p^{-1}(k) = i \\
	 \mathcal{A}(k) \pm 1 & i < p^{-1}(k) \leq j \\
	 \mathcal{A}(k) & \text{ otherwise }
\end{cases}
\]
Thus, $\mathcal{A}$ may be partitioned into at most four contiguous intervals, each upon which the update is constant. Such updates are called \emph{range updates}, and are known to require $O(\log m)$ time using e.g. an \emph{implicit treap} data structure~\cite{blelloch1998fast}. 
Since single element modifications, deletions, insertions, and constant range updates can all be achieved in $O(\log m)$ expected time with such a data structure, we conclude that equation~\eqref{eq:greedy_step} may be solved in just $O(d^2 \log m)$ time. 
We give an example of this partitioning in Figure~\ref{fig:displacement_proof}.

\section{Applications and Experiments}\label{sec:results}
\subsection{Video data}
A common application of persistence is characterizing topological structure in image data. Since a  set of ``snapshot'' frames of a video can be equivalently thought of as discrete 1-parameter family, our framework provides a natural extension of the typical image analysis to video data. 
To demonstrate the benefit of scheduling and the scalability of the greedy heuristic, we perform two performance tests on the video data from section~\ref{sec:motivation}: one to test the impact of repairing the decomposition less and one to measure the asymptotic behavior of the greedy approach.

\begin{figure}
	\centering
	\includegraphics[width=0.48\textwidth]{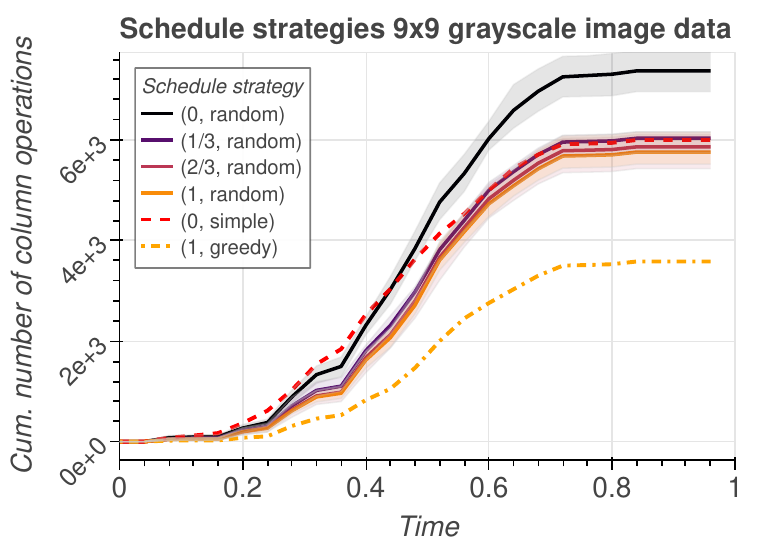}
	\includegraphics[width=0.48\textwidth]{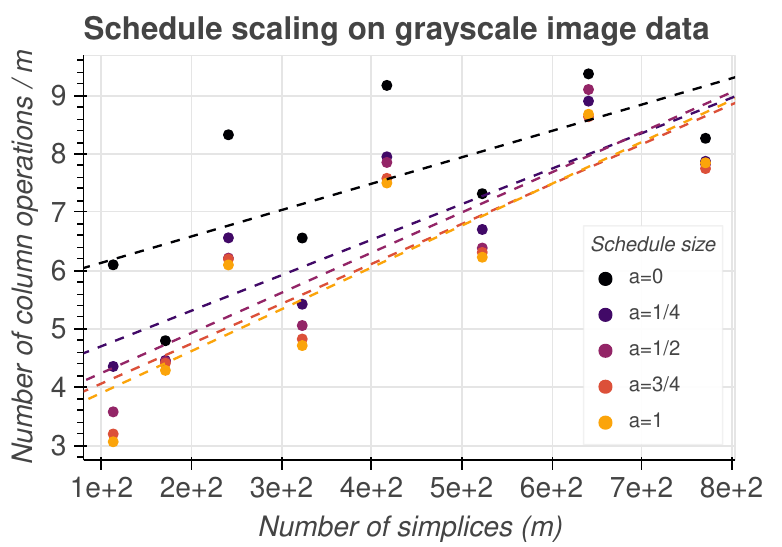}
	\caption{Performance comparison between various scheduling strategies. On the left, the cumulative column operations required to compute the 1-parameter family is shown for varying schedule sizes ($d$) and strategies. On the right, both the size of the schedule and the data set $(m)$ are varied. }
	\label{fig:movie_perf}
\end{figure}

 In the first test, we fix a grid size of $9 \times 9$ and record the cumulative number of column operations needed to compute persistence dynamically across 25 evenly-spaced time points using a variety of scheduling strategies.
 The three strategies we test are the greedy approach from section~\ref{sec:proxy_objective}, the ``simple'' approach which uses upwards of $O(m)$ move permutations via selection sort, and a third strategy which interpolates between the two. 
To perform this interpolation, we use a parameter $\alpha \in [0,1]$ to choose $m - \alpha \cdot d$ random simplices to move using the same construction method outlined in section~\ref{sec:schedule_construction}. 
The results are summarized in the left graph on Figure~\ref{fig:movie_perf}, wherein the mean schedule cost of the random strategies are depicted by solid lines. To capture the variation in performance, we run 10 independent iterations and shade the upper and lower bounds of the schedule costs. 
 As seen in Figure~\ref{fig:movie_perf}, while using less move operations (lower $\alpha$) tends to reduce column operations, constructing \emph{random} schedules of minimal size is no more competitive than the selection sort strategy. This suggests that efficient schedule construction needs to account for the structure of performing several permutations in sequence, like the greedy heuristic we introduced, to yield an adequate performance boost. 

In the second test, we measure the asymptotics of our greedy LCS-based approach. To do this, we generated 8 video data sets again of the expanding annulus outlined in section~\ref{sec:motivation}, each of increasing grid sizes of $5 \times 5$, $6 \times 6$, $\dots$, $12 \times 12$. For each data set, we compute persistence over the duration of the video, again testing five evenly spaced settings of $\alpha \in [0,1]$---the results are shown in the right plot of Figure~\ref{fig:movie_perf}. On the vertical axis, we plot the total number of column operations needed to compute persistence across 25 evenly-spaced time points \emph{as a ratio} of the data set size ($m$); we also show the regression curves one obtains for each setting of $\alpha$. As one can see from the Figure, the cost of using the greedy heuristic tends to increase sub-linearly as a function of the data set size, suggesting the move scheduling approach is indeed quite scalable. Moreover, schedules with minimal size tended to be cheaper than otherwise, confirming our initial hypothesis that repairing the decomposition less can lead to substantial reductions at run-time.  

\subsection{Crocker stacks}
There are many challenges to characterizing topological behavior in dynamic settings. One approach is to trace out the curves constituting a continuous family of persistence diagrams in $\mathbb{R}^3$---the \emph{vineyards} approach---however this visualization can be cumbersome to work with as there are potentially many such vines tangled together, making topological critical events with low persistence difficult to detect. Moreover, the \emph{vineyards} visualization does not admit a natural simplification utilizing the stability properties of persistence, as individual vines are not stable: if two vines move near each other and then pull apart without touching, then a pairing in their corresponding persistence diagrams may cross under a small perturbation, signaling the presence of an erroneous topological critical event~\cite{topaz2015topological, xian2020capturing}. 

\begin{figure}[t]
	\centering
	\includegraphics[width=0.95\textwidth]{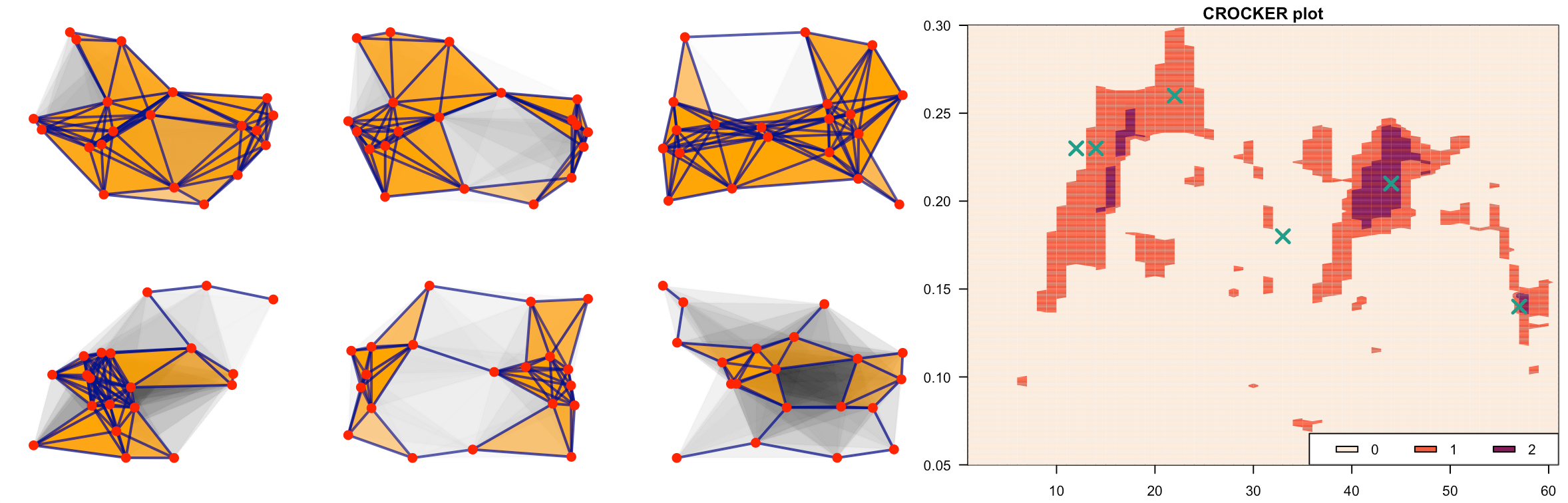}
	\caption{A Crocker plot (right) depicts the evolution of dimension $p = 1$ Betti curves over time. The green $\mathrm{X}$ marks correspond chronologically to the complexes (left), in row-major order. The large orange and purple areas depict $1$-cycles persisting in both space (y-axis) and time (x-axis).} \label{fig:crocker1}
\end{figure}

Acknowledging this, Topaz et al.~\cite{topaz2015topological} proposed the use of a 2-dimensional summary visualization, called a \emph{crocker}\footnote{\emph{crocker} stands for ``Contour Realization Of Computed k-dimensional hole Evolution in the Rips complex.'' Although the acronym includes \emph{Rips complexes} in the name, in principle a \emph{crocker} plot could just as easily be created using other types of triangulations (e.g. \v{C}ech filtrations).} plot. 
In brief, a \emph{crocker} plot is a contour plot of a family of Betti curves. Formally, given a filtration $K = K_0 \subseteq K_1 \subseteq \dots \subseteq K_m$, a $p$-dimensional \emph{Betti curve} $\beta_p^{\bullet}$ is defined as the ordered sequence of $p$-th dimensional Betti numbers:
$$ \beta_p^\bullet = \{ \, \mathrm{rank}(H_p(K_0)), \, \mathrm{rank}(H_p(K_1)), \, \dots, \, \mathrm{rank}(H_p(K_m))\, \}$$
Given a time-varying filtration $K(\tau)$, a \emph{crocker} plot displays changes to $\beta_p^\bullet(\tau)$ as a function of $\tau$. An example of a \emph{crocker} plot generated from the simulation described below is given in Figure~\ref{fig:crocker1}. Since only the Betti numbers at each simplex in the filtration are needed to generate these Betti curves, the persistence diagram is not directly needed to generate a \emph{crocker} plot; it is sufficient to use e.g. any of the specialized methods discussed in~\ref{sec:related_work}. This dependence only on the Betti numbers makes \emph{crocker} plots easier to compute than standard persistence, however what one gains in efficiency one loses in stability; it is known that Betti curves are inherently unstable with respect to small fluctuations about the diagonal of the persistence diagram. 

Xian et al.~\cite{xian2020capturing} showed that \emph{crocker} plots may be \emph{smoothed} to inherit the stability property of persistence diagrams and reduce noise in the visualization. That is, when applied to a time-varying persistence module $M = \{M_t\}_{t \in [0, T]}$, an $\alpha$-smoothed \emph{crocker} plot for $\alpha \geq 0$ is the rank of the map $M_t(\epsilon - \alpha) \to M_t(\epsilon + \alpha)$ at time $t$ and scale $\epsilon$. For example, the standard crock plot is a $0$-smoothed \emph{crocker} plot. Allowing all three parameters ($t, \epsilon, \alpha$) to vary continuously leads to 3D visualization called an $\alpha$\emph{-smoothed crocker stack}.
\begin{definition}[crocker stack]
	A \emph{crocker stack} is a family of $\alpha$-smoothed crock plots which summarizes the topological information of a time-varying persistence module $M$ via the function $f_M : [0, T] \times [0, \infty) \times [0, \infty) \to \mathbb{N}$, where:
	$$ f_M(t,\epsilon, \alpha) = \mathrm{rank}(M_t(\epsilon - \alpha) \to M_t(\epsilon + \alpha)) $$
	and $f_M$ satisfies $f_M(t,\epsilon,\alpha') \leq f_M(t,\epsilon, \alpha)$ for all $0 \leq \alpha \leq \alpha'$.
\end{definition}
\noindent Note that, unlike \emph{crocker} plots, applying this $\alpha$ smoothing efficiently \emph{requires} the persistence pairing. Indeed, it has been shown that \emph{crocker stacks} and stacked persistence diagrams (i.e. \emph{vineyards}) are equivalent to each other in the sense that either one contains the information needed to reconstruct the other~\cite{xian2020capturing}. Thus, computing \emph{crocker stacks} reduces to computing the persistence of a (time-varying) family of filtrations.


To illustrate the applicability of our method, we test the efficiency of computing these \emph{crocker stacks} using a spatio-temporal data set. Specifically, we ran a flocking simulation similar to the one run in~\cite{topaz2015topological} with $m = 20$ vertices moving around on the unit square equipped with periodic boundary conditions (i.e. $S^1 \times S^1$). We simulated movement by equipping the vertices with a simple set of rules which control how the individual vertices position change over time. Such simulations are also called \emph{boid} simulations, and they have been extensively used as models to describe how the evolution of collective behavior over time can be described by simple sets of rules.
The simulation is initialized with every vertex positioned randomly in the space; the positions of vertices over time is updated according to a set of rules related to the vertices' acceleration, distance to other vertices, etc. To get a sense of the time domain, we ran the simulation until a vertex made at least 5 rotations around the torus. 

\begin{figure}[ht]
	\centering
	\includegraphics[width=0.6\textwidth, height=14.15em]{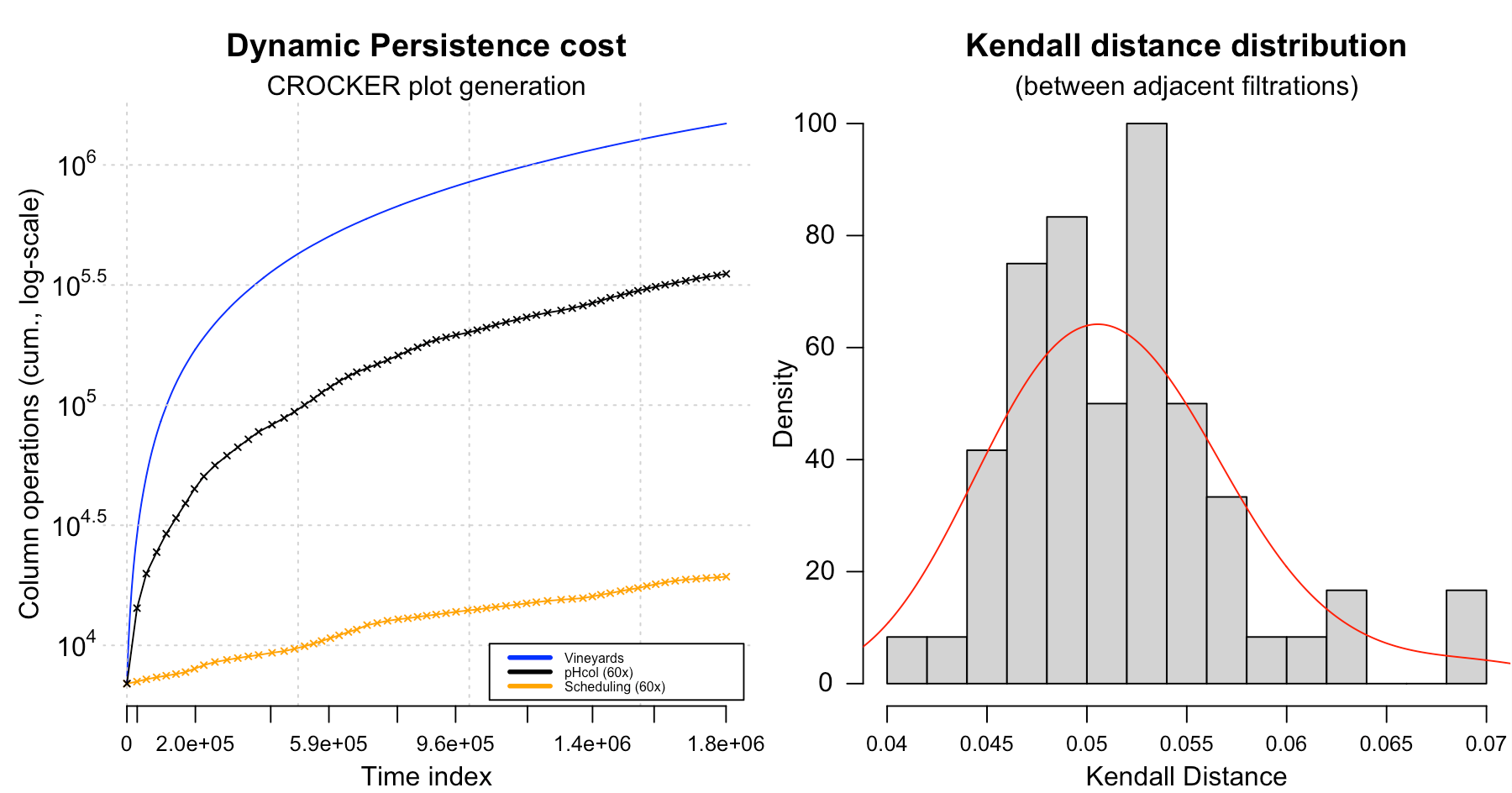} 
	\caption{On the left, the cumulative number of column operations (log-scale) of the three baseline approaches tested. On the right, the normalized $K_\tau$ between adjacent filtrations depicts the coarseness of the discretization---about $5\%$ of the $\approx O(m^2)$ simplex pairs between adjacent filtrations are discordant.}
	\label{fig:boid_sim_results}
\end{figure}

Given this time-evolving data set, we computed the persistence diagram of the Rips filtration up to $\epsilon = 0.30$ at 60 evenly spaced time points using three approaches: the standard algorithm \texttt{pHcol} applied naively at each of the 60 time steps, the \emph{vineyards} algorithm applied to (linear) homotopy connecting filtrations adjacent in time, and our approach using \emph{moves}.   
The cumulative number of $O(m)$ column operations executed by three different approaches. Note again that \emph{vineyards} requires generating many decompositions by design (in this case, $\approx 1.8M$). The standard algorithm \texttt{pHcol} and our move strategy were computed at 60 evenly spaced time points. As depicted in Figure~\ref{fig:boid_sim_results}, our \emph{moves} strategy is far more efficient than both \emph{vineyards} and the naive \texttt{pHcol} strategies. 

\subsection{Multiparameter persistence}
Given a procedure to filter a space in multiple dimensions simultaneously, a \emph{multifiltration}, the goal of multi-parameter persistence is to identify persistent features by examining the entire multifiltration. 
Such a generalization has appeared naturally in many application contexts, showing potential as a tool for exploratory data analysis~\cite{lesnick2012multidimensional}. Indeed, one of the drawbacks of persistence is its instability with respect to strong outliers, which can obscure the detection of significant topological structures~\cite{buchet2015topological}.
One exemplary use case of multi-parameter persistence is to detect these strong outliers by filtering the data with respect to both the original filter function \emph{and} density.
In this section, we show the utility of scheduling with a real-world use case: detecting the presence of a low-dimensional topological space which well-approximates the distribution of natural images. 
As a quick outline, in what follows we briefly recall the fibered barcode invariant~\ref{sec:fibered_barcode}, summarize its potential application to a particular data set with known topological structure~\ref{sec:natural_images}, and conclude with experiments of demonstrating how scheduling enables such applications~\ref{sec:empirical_klein}.   

\subsubsection{Fibered barcode}\label{sec:fibered_barcode}
Unfortunately, unlike the one-parameter case, there is no complete discrete invariant for multi-parameter persistence.
Circumventing this, Lesnick et al~\cite{lesnick2015interactive} associate a variety of incomplete invariants to 2-parameter persistence modules; we focus here on the \emph{fibered barcode} invariant, defined as follows: 
\begin{definition}[Fibered barcode]
	The fibered barcode $\mathcal{B}(M)$ of a 2D persistence module $M$ is the map which sends each line  $L \subset \mathbb{R}^2$ with non-negative slope to the barcode $\mathcal{B}_L(M)$: 
$$ \mathcal{B}(M) = \{ \; B_L(M) : L \in \mathbb{R} \times \mathbb{R}^{+} \; \}$$
Equivalently, $\mathcal{B}(M)$ is the 2-parameter family of barcodes given by restricting $M$ to the of set affine lines with non-negative slope in $\mathbb{R}^2$. 
\end{definition} 
\noindent Although an intuitive invariant, it is not clear how one might go about computing $\mathcal{B}(M)$ efficiently. 
One obvious choice is fix $L$ via a linear combination of two filter functions, restrict $M$ to $L$, and compute the associated 1-parameter barcode. 
However, this is an $O(m^3)$ time computation, which is prohibitive for interactive data analysis purposes. 

\begin{figure}[t]
	\includegraphics[width=0.98\textwidth]{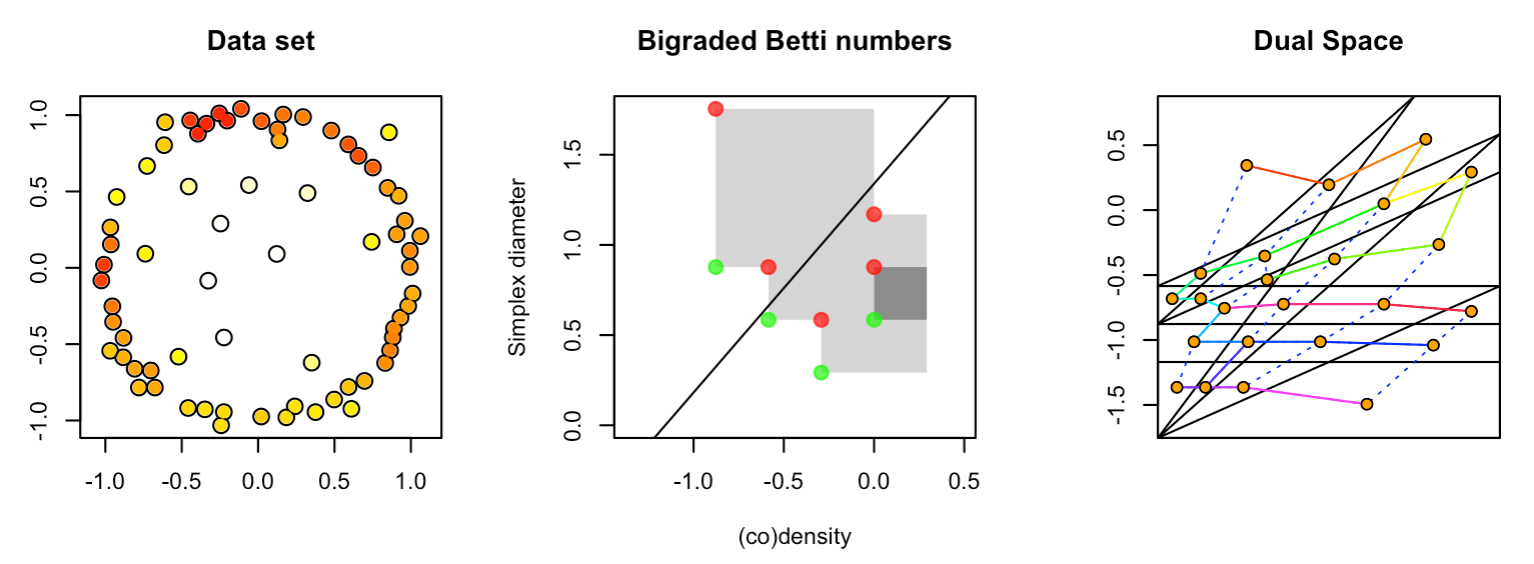}
	\caption{Bipersistence example on an $8 \times 8$ coarsened grid. On the left, the input data, colored by density. In the middle, the bigraded Betti numbers $\beta_0(M)$ and $\beta_1(M)$ (green and red, respectively), the dimension function (gray), and a line $L$ emphasizing the persistence of features with high density. On the right, the line arrangement $\mathcal{A}(M)$ lying in the dual space derived from the $\beta(M)$. }
	\label{fig:bifiltration_ex2}
\end{figure}

Utilizing the equivalence between the rank and fibered barcode invariants, Lesnick and Wright~\cite{lesnick2015interactive} developed an elegant way of computing $\mathcal{B}(M)$ via a re-parameterization using standard point-line duality. This clever technique effectively reduces the fibered barcode computation to a sequence of 1-D barcode computations at ``template points'' lying within the 2-cells of a particular planar subdivision $\mathcal{A}(M)$ of the half-plane $[0, \infty) \times \mathbb{R}$. This particular subdivision is induced by the arrangement of ``critical lines'' derived by the bigraded Betti numbers $\beta(M)$ of $M$.
As the barcode of one template point $\mathcal{T}_e$ at the 2-cell $e \in \mathcal{A}(M)$ may be computed efficiently by re-using information from an adjacent template point $\mathcal{T}_{e'}$,~\cite{lesnick2015interactive} observed that computing the barcodes of all such template points (and thus, $\mathcal{B}(M)$) may be reduced to ordering the 2-cells in $\mathcal{A}(M)$ along an Eulerian path traversing the dual graph of $\mathcal{A}(M)$. 
The full algorithm is out of scope for this effort; we include supplementary details for the curious reader in the appendix~\ref{app:2d_pers}. 
\\
\\
\noindent 
\textbf{Example 4.1}: Consider a small set of noisy points distributed around $S^1$ containing a few strong outliers, as shown on the left side of Figure~\ref{fig:bifiltration_ex2}. Filtering this data set with respect to the Rips parameter and the complement of a kernel density estimate yields a bifiltration whose various invariants are shown in the middle figure. The gray areas indicate homology with positive dimension---the lighter gray area $\mathrm{dim}_1(M) = 1$ indicates a persistent loop was detected.
On the right side, dual space is shown: the black lines are the critical lines that form $\mathcal{A}(M)$, the blue dashed-lines the edges of the dual graph of $\mathcal{A}(M)$, the rainbow lines overlaying the dashed-lines form the Eulerian path, and the orange barycentric points along the 2-cells of $\mathcal{A}(M)$ represent where the barcodes templates $\mathcal{T}_e$ are parameterized. 
\\

Despite its elegance, there are significant computational barriers prohibiting the 2-parameter persistence algorithm from being practical. 
An analysis from~\cite{lesnick2015interactive} (using \emph{vineyards}) shows the barcodes template computation requires on the order of $O(m^3 \kappa + m \kappa^2 \log \kappa)$ elementary operations and $O(m \kappa^2)$ storage, where $\kappa$ is a coarseness parameter.
Since the number of 2-cells in $\mathcal{A}(M)$ is on the order $O(\kappa^2)$, and $\kappa$ itself is on the order of $O(m^2)$ in the worst case, the scaling of the barcode template computation may approach $\approx O(m^5)$---this is both the highest complexity and most time-intensive sub-procedure the RIVET software~\cite{rivet} depends on. 
Despite this significant complexity barrier, in practice the external stability result from~\cite{landi2014rank} justifies the use of a grid-like reduction procedure which approximates the module $M$ with a smaller module $M'$, enabling practitioners to restrict the size of $\kappa$ to a relatively small constant. 
This in-turn dramatically reduces the size of $\mathcal{A}(M)$ and thus the number of barcode templates to compute. 
Moreover, the ordering of barcode templates given by the dual graph traversal implies that adjacent  template points should be relatively close---so long as $\kappa$ is not too small---suggesting adjacent templates may productively share computations due to the high similarity of their associated filtrations. 
Indeed, as algorithm~\ref{alg:schedule} was designed for precisely such a computation, 2-parameter persistence is prototypical of the class of methods that stand to benefit from \emph{moves}. 

 \subsubsection{Natural images dataset}\label{sec:natural_images}
A common hypothesis is that high dimensional data tend to lie in the vicinity of an embedded, low dimensional manifold or topological space. An exemplary demonstration of this is given in the analysis by Lee et al.~\cite{lee2003nonlinear}, who explored the space of high-contrast patches extracted from Hans van Hateren's~\cite{hateren_schaaf_1998} still image collection\footnote{See \url{http://bethgelab.org/datasets/vanhateren/} for details on the image collection.}, which consists of $\approx 4,000$ monochrome images depicting various areas outside Groningen (Holland). 
In particular,~\cite{lee2003nonlinear} were interested in exploring how high-contrast $3 \times 3$ image patches  were distributed, in pixel-space, with respect to predicted spaces and manifolds.
Formally, they measured contrast using a discrete version of the scale-invariant Dirichlet semi-norm:
$$ \lVert x \rVert_D = \sqrt{\sum_{i \sim j}(x_i - x_j)^2} = \sqrt{x^T D x}$$
where $D$ is a fixed matrix whose quadratic form $x^T D x$ applied to an image $x \in \mathbb{R}^9$ is proportional to the sum of the differences between each pixels 4 connected neighbors (given above by the relation $i \sim j$).
Their research was primarily motivated by discerning whether there existed clear qualitative differences in the distributions of patches extracted from images of different modalities, such as optical and range images.
By mean-centering, contrast normalizing, and``whitening'' the data via the Discrete Cosine Transform (DCT), they show a convenient basis for $D$ may be obtained via an expansion of 8 certain non-constant eigenvectors, shown below: 
\begin{center}
	\includegraphics[width=0.80\textwidth]{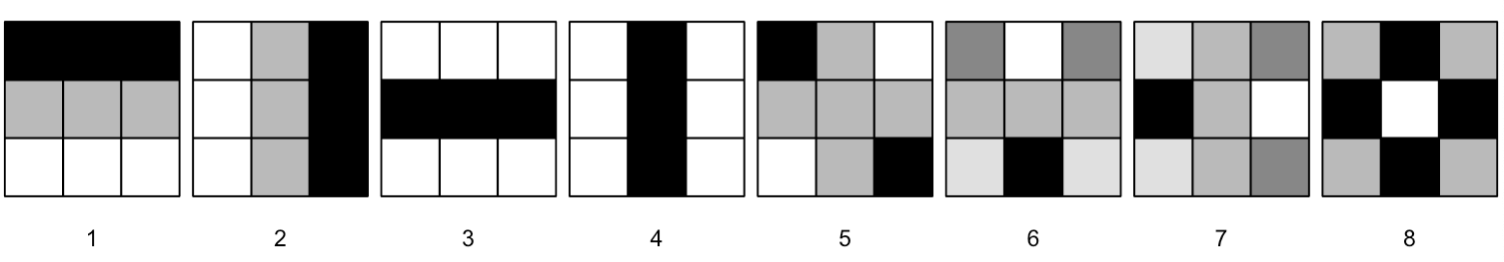} 
\end{center}
Since these images are scale-invariant, the expansion of these basis vectors spans the 7-sphere, $S^7 \subset \mathbb{R}^8$. Using a Voronoi cell decomposition of the data, their distribution analysis suggested that the majority of data points concentrated in a few high-density regions. 

In follow-up work, Carlsson et al.~\cite{carlsson2008local} found---using persistent homology---that the distribution of high-contrast $3 \times 3$ patches is actually well-approximated by a Klein bottle $\mathcal{M}$---around 60\% of the high-contrast patches from the still image data set lie within a small neighborhood around $\mathcal{M}$ accounting for only 21\% of the 7-sphere's volume. 
Along a similar vein, Perea~\cite{perea2014klein} established a dictionary learning framework for efficiently estimating the distribution of patches from texture images, prompting applications for persistent homology in sparse coding contexts. 

 \begin{figure}[t]
	\includegraphics[width=0.98\textwidth]{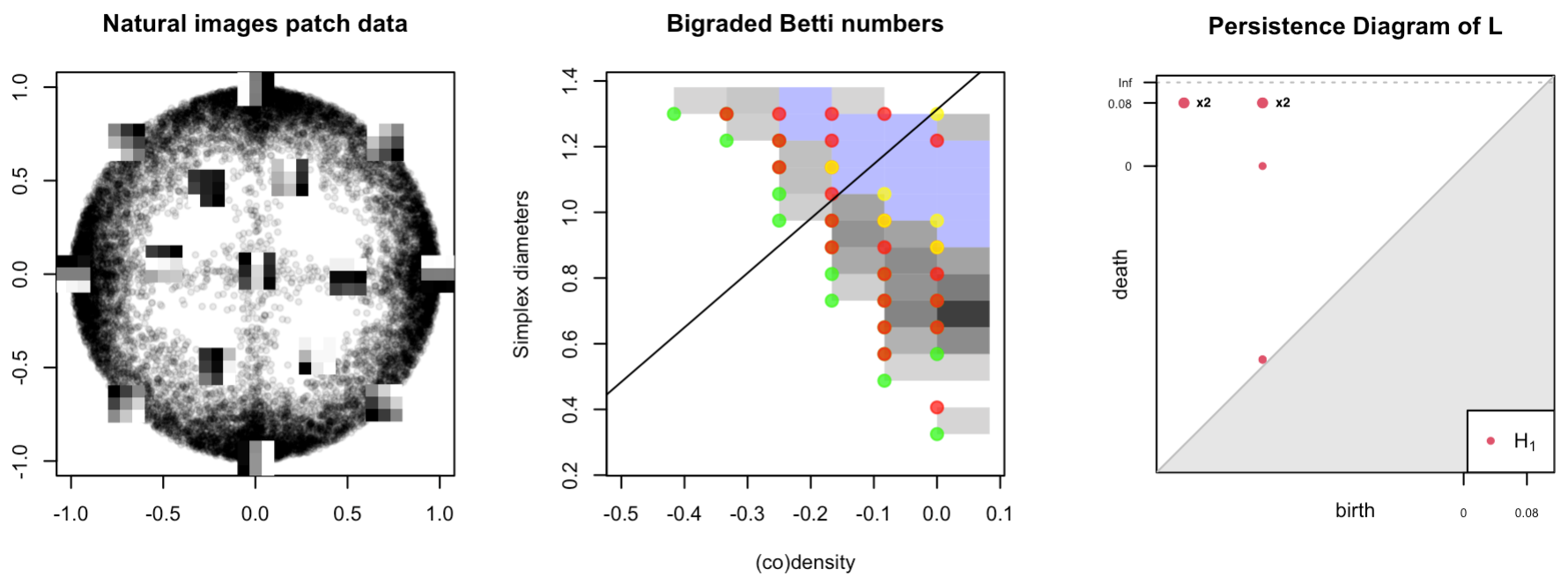}
	\caption{Bipersistence example of natural images data set on a $12 \times 16$ coarsened grid. On the left, a projection of the full data set is shown, along with the 15 landmark patches. (Middle) the bigraded Betti numbers and a fixed line $L$ over parameter space. 
	As before, the $0/1/2$ dimension bigraded Betti numbers are shown in green/red/yellow, respectively, with the blue region highlighting where $\mathrm{dim}(M) = 5$. (Right) five persistent features representing $B_L(M)$ are revealed from the middle, matching $\beta_1$ of the three-circle model. }
	\label{fig:patch_data_dgm}
\end{figure}

If one was not aware of the analysis done by~\cite{lee2003nonlinear, hateren_schaaf_1998, carlsson2008local, perea2014klein}, it is not immediately clear a priori that the Klein bottle model is  a good candidate for capturing the non-linearity of image patches. 
Indeed, armed with a refined topological intuition, Carlsson still needed to perform extensive sampling, preprocessing, and model fitting techniques in order to reveal the underlying the topological space with persistent homology~\cite{carlsson2008local}.
One reason such preprocessing is needed is due to persistent homology's aforementioned instability with respect to strong outliers. 
In the ideal setting, a multi-parameter approach that accounts for the local density of points should require far less experimentation. 
 
To demonstrate the benefit of 2-parameter persistence on the patch data, consider the (coarsened) fibered barcode computed from a standard Rips / codensity bifiltration on a representative sample of the image data from~\cite{hateren_schaaf_1998}, shown in Figure~\ref{fig:patch_data_dgm}. 
From the bigraded Betti number and the dimension function, one finds that a large area of the dimension function is constant (highlighted as the blue portion in the middle of Figure~\ref{fig:patch_data_dgm}), wherein the first Betti number is 5. Further inspection suggests one plausible candidate is the three-circle model $C_3$, which consists of three circles, two of which (say, $S_v$ and $S_h$) intersect the third (say, $S_\mathrm{lin}$) in exactly two points, but themselves do not intersect. 
Projecting the image data onto the first two basis vectors from the DCT shown above leads to the projection shown in the top left of Figure~\ref{fig:patch_data_dgm}, of which 15 landmark points are also shown. Observe the data are distributed well around three ``circles''---the outside circle capturing the rotation gradient of the image patches ($S_{\mathrm{lin}}$), and the other two capturing the vertical and horizontal gradients ($S_v$ and $S_h$, respectively). Since the three circle model is the 1-skeleton of the Klein bottle, one may concur with Carlssons analysis~\cite{carlsson2008local} that the Klein bottle may be a reasonable candidate upon which the image data are distributed. 

The degree to which multi-parameter persistence simplifies this exploratory phase cannot be understated: we believe multi-parameter persistence has a larger role to play in manifold learning. 
Unfortunately, as mentioned prior, the compute barriers effectively bar its use in practice. 

\subsubsection{Accelerating 2D persistence}\label{sec:empirical_klein}
Having outlined the computational theory of 2-parameter persistence, we now demonstrate the efficiency of \emph{moves} using the same high-contrast patch data set studied in~\cite{lee2003nonlinear} by evaluating the performance of various methods at computing the fibered barcode invariant via the parameterization from~\ref{sec:fibered_barcode_reparam}. 

Due to the aforementioned high complexity of the fibered barcode computation, we begin by working with a subset of the image patch data $\mathcal{X}$. 
We combine the use of furthest-point sampling and proportionate allocation (stratified) sampling to sample landmarks $X \subset \mathcal{X}$ distributed within $n = 25$ strata. Each stratum consists of the $(1/n)$-thick level set given the $k$-nearest neighbor density estimator $\rho_{15}$ with $k = 15$. The use of furthest-point sampling gives us certain coverage guarantees that the geometry is approximately preserved within each level set, whereas the stratification ensures the original density of is approximated preserved as well. 
From this data set, we construct a Rips-(co)density bifiltration using $\rho_{15}$ equipped with the geodesic metric computed over the same k-nearest neighbor graph on $X$. 
Finally, we record the number of column reductions needed to compute the fibered barcode at a variety of levels of coarsening using $\mathtt{pHcol}$, \emph{vineyards}, and our \emph{moves} approach. The results are summarized in Table~\ref{tab:barcode_templates}. We also record the number of 2-cells in $\mathcal{A}(M)$ and the number of permutations applied encountered along the traversal of the dual graph for both \emph{vineyards} and \emph{moves}, denoted in the table as $d_K$ and $d_{\mathrm{LCS}}$, respectively. 
\begin{table}[h]
\caption{Cost to computing $\mathcal{T}$ for various coarsening choices of $\beta(M)$.}\label{tab:barcode_templates}
\centering
\begin{tabular}{ m{1.6cm} m{1.6cm} m{1.2cm} m{2.8cm} m{2.4cm} }
 \hline
  \multicolumn{1}{|c|}{ $\beta(M)$ } & \multicolumn{1}{|c|}{$\mathcal{A}(M)$ } & \multicolumn{3}{c|}{ Col. Reductions / Permutations } \\
 \hline
\small{Coarsening} & \# 2-cells & \texttt{phCol} & Vineyards / $d_{K}$ & Moves / $d_{\text{LCS}}$ \\
 \hline
 8 x 8 & 39 & 94.9K & 245K / 1.53M & 38.0K / 11.6K   \\
 \hline
 12 x 12 & 127 & 318K & 439K / 2.66M & 81.9K / 33.0K \\
 \hline 
 16 x 16 & 425 & 1.07M & 825K / 4.75M & 114K / 87.4K \\
 \hline 
  20 x 20 & 926 & 2.32M & 1.15M / 6.77M  & 148K / 154K\\
 \hline 
  24 x 24 & 1.53K & 3.92M & 1.50M / 8.70M & 184K / 232K  \\
   \hline
 \end{tabular}
\label{table:dual_graph_costs}
\end{table}  

As shown on the table, when the coarsening $\kappa$ is small enough, we're able to achieve a significant reduction in the number of total column operations needed to compute $\mathcal{T}$ compared to both \emph{vineyards} and $\mathtt{pHcol}$. 
This is further reinforced by the observation that \emph{vineyards} is particularly inefficient when then 1-parameter family is coarse. Indeed, \emph{moves} requires about 3x less column operations than naively computing $\mathcal{T}$ independently. However, note that as the coarsening becomes more refined and more 2-cells are added to $\mathcal{A}(M)$, \emph{vineyards} becomes a more viable option compared to $\mathtt{pHcol}$---as the asymptotics suggests---though even at the highest coarsening we tested the gain in efficiency is relatively small. In contrast, \emph{moves} scales quite well with this refinement, requiring about $12\%$  and $\approx 5\%$ of the number of column operations as \emph{vineyards} and $\mathtt{pHcol}$, respectively.




\section{Conclusion and Future Work}\label{sec:conclusion}
In conclusion, we presented a scheduling algorithm for efficiently updating a decomposition in coarse dynamic settings. Our approach is simple, relatively easy to implement, and fully general: it does not depend on the geometry of underlying space, the choice of triangulation, or the choice of homology dimension. Moreover, we supplied efficient algorithms for our scheduling strategy, provided tight bounds where applicable, and demonstrated our algorithms performance with several real world use cases.

There are many possible applications of our work beyond the ones discussed in section~\ref{sec:results}, such as e.g. accelerating PH featurization methods or detecting homological critical points in dynamic settings.
Indeed, we see our approach as potentially useful to any situation where the structure of interest can be cast as a parameterized family of persistence diagrams. Areas of particular interest include time-series analysis and dynamic metric spaces~\cite{kim2020spatiotemporal}. 

The simple and combinatorial nature of our approach does pose some limitations to its applicability. For example, better bounds or algorithms may be obtainable if stronger assumptions can be made on how the filtration is changing with time. Moreover, if the filtration $(K, f)$ shares little similarity to the ``target'' filtration $(L,f')$, then the overhead of reducing the simplices from $L \setminus K$ appended to the decomposition derived from $K$ may be large enough to motivate simply computing the decomposition at $L$ independently. 
Our approach is primarily useful if the filtrations in the parameterized family are ``nearby'' in the combinatorial sense. 

From an implementation perspective, one non-trivial complication of our approach is its heavy dependence on a particular sparse matrix data structure, which permits permuting both the row and columns of a given matrix in at most $O(m)$ time~\cite{cohen2006vines}. As shown with the natural images example in section~\ref{sec:results}, there are often more permutation operations being applied than there are column reductions. In the more standard \emph{compressed} sparse matrix representations,
permuting both the rows and columns generally takes at most $O(Z)$ time, where $Z$ is the number of non-zero entries, which can be quite expensive if the particular filtration has many cycles. As a result, the more complex sparse matrix representation from~\cite{cohen2006vines} is necessary to be efficient in practice. 

Moving forward, our results suggest there are many aspects of computing persistence in dynamic settings yet to be explored. 
For example, it's not immediately clear whether one could adopt, for example, the twist optimization~\cite{chen2011persistent} used in the reduction algorithm to the dynamic setting. Another direction to explore would be the analysis of our approach under the cohomology computation~\cite{de2011dualities}, or the specialization of the move operations to specific types of filtrations such as Rips filtrations. Such adaptations may result in even greater reductions in the number of column operations, as have been observed in practice for the standard reduction algorithm~\cite{bauer2021ripser}. 
Moreover, though we have carefully constructed an efficient greedy heuristic in section~\ref{sec:proxy_objective} and illustrated a different perspective with which to view our heuristic (via crossing minimization), it is an open question whether there exists a more structured reduction of~\eqref{eq:schedule_cost} or~\eqref{eq:sorting_additive} to a better-known problem.

\section{Declarations}
 
\textbf{Ethical Approval:} Not applicable.
\\
\\
\noindent \textbf{Competing interests:} 
The authors declare that they have no competing interests that could influence the interpretation or presentation of the research findings. There are no financial or personal relationships with individuals or organizations that could bias the outcomes of this work.
\\
\\
\noindent
\textbf{Authors' contributions:}
The contributions of each author to this article were as follows:
\begin{enumerate}
	\item Matt Piekenbrock: conceptualization, methodology design, algorithm development, experimental design \& analysis, software development. 
	\item Jose Perea: Conceptualization, methodology design, literature review, critical revision of the manuscript, final approval of the version to be published.
\end{enumerate}
All authors have reviewed and approved the final version of the manuscript and have agreed to be accountable for all aspects of the work.
\\
\\
\noindent 
\textbf{Funding:}
The research presented in this work is partially supported by the National Science Foundation through grants CCF-2006661 and CAREER award DMS-1943758. The funding source had no role in the study design, data collection and analysis, decision to publish, or preparation of the manuscript.
\\
\\
\noindent \textbf{Availability of data and materials:} Links to the software and data used for the experiments can be found at: \url{https://github.com/peekxc/move_schedules}.

\bibliography{move_schedules}

\newpage 
\appendix
\section{Appendix}

\subsection{Algorithms}

\textbf{Reduction Algorithm: } The reduction algorithm, outlined in pseudocode in Algorithm~\ref{alg:reduce}, is the most widely used modality for computing persistence. While there exists other algorithms for computing persistence, they are typically not competitive with the reduction algorithm in practice. 
\begin{algorithm}[h!]
	\setstretch{1.15}
	\caption{Reduction Algorithm (\texttt{pHcol}) }
	\begin{algorithmic}[1]
		\Require{$D = (m \times m)$ filtration boundary matrix }
		\Ensure{$R$ is reduced, $V$ is full rank upper triangular, and $R = D V$}
		\Function{Reduction}{$D$}
		\State $(R, V) \gets (D, I)$
		\For{$j = 1$ \textbf{to} $m$} 
			\While{$\exists \, i < j$ \textbf{such that} $\mathrm{low}_R(i) = \mathrm{low}_R(j)$ }
				\State $\lambda \gets \mathrm{pivot}_R(j)/\mathrm{pivot}_R(i)$
				\State $(\mathrm{col}_R(j), \mathrm{col}_V(j)) \mathrel{-}= \left ( \lambda \cdot \mathrm{col}_R(i), \lambda \cdot \mathrm{col}_V(i) \right )$
			\EndWhile
		\EndFor 
		\State \Return $(R, V)$
		\EndFunction
	\end{algorithmic}	\label{alg:reduce}
\end{algorithm}
The algorithm begins by copying $D$ to a new matrix $R$, to be subsequently modified in-place. After  $V$ is set to the identity, the algorithm proceeds with column operations on both $R$ and $V$, left to right, until the decomposition invariants are satisfied. Since each column operation takes $O(m)$ and there are potentially $O(k)$ columns in $D$ with identical low entries (line 4 in ~\ref{alg:reduce}, observe that the reduction algorithm below clearly takes $O(m^2 k)$ time. Since there exist complexes where $k \sim O(m)$, one concludes the bound of $O(m^3)$ is tight~\cite{morozov2005persistence}, though this seems to only be true on pathological inputs. A more refined analysis by Edelsbrunner et al.~\cite{edelsbrunner2000topological} shows the reduction algorithm scales by the sum of squares of the cycle persistences. 
\\
\\
\noindent \textbf{Move Left:}
Though conceptually similar, note that there is an asymmetry between \emph{MoveRight} and \emph{MoveLeft}: moving a simplex upwards in the filtration requires removing non-zero entries along several columns of a particular row in $V$ so that the corresponding permutation does not render $V$ non-upper triangular. The key insight of the algorithm presented in~\cite{busaryev2010tracking} is that $R$ can actually be maintained in all but one column during this procedure (by employing the \emph{donor} column). In contrast, moving a simplex to an earlier time in the filtration requires removing non-zero entries along several rows of a particular column of $V$. As before, though $R$ stays reduced during this cancellation procedure in all but one column, the subsequent permutation to $R$ requires reducing a pair of columns which may cascade into a larger chain of column operations to keep $R$ reduced. 
This is due to the fact that higher entries in columns in $R$ (above the pivot entry) may very well introduce additional non-reduced columns after $R$ is permuted. 
Since these operations always occur in a left-to-right fashion, it is not immediately clear how to apply a donor column kind of concept. Fortunately, like move right, we can still separate the algorithm into a reduction and restoration phase---see Algorithm~\ref{alg:ml}. Moreover, since $R$ is reduced in all but one column by line 6 in~Algorithm~\ref{alg:ml}, we can still guarantee the number of low entries to reduce in $R$ will be at most $\lvert i - j \rvert$. For a supplementary description of the move algorithm, see~\cite{busaryev2010tracking}.
\\
\\
\noindent \textbf{LCS-Sort:}
The algorithm to construct a schedule from a given LIS is simple enough to derive using the rules discussed in section~\ref{sec:schedule_construction} (namely, equation~\eqref{eq:valid_move}, but nonetheless for posterity’s sake we record it here for the curious reader; the pseudocode is given in Algorithm~\ref{alg:sorting}.

The high level idea of the algorithm is to first construct the LCS between two permutations $p,q \in S_m$ and then apply cyclic permutations successively to $p$, each step adding a new element extending the LCS. 
The algorithmic steps are as follows: first, one re-labels $q \mapsto \iota$ to the identity permutation $\iota = [m]$ and applies a consistent re-labeling $p \mapsto \bar{p}$. This relabeling preserves the LCS distance and has the additional advantage that $\bar{q} = \iota = [m]$ is a strictly increasing subsequence, and thus computing the LCS between $p,q \in S_m$ reduces to computing the LIS $\mathcal{L}$ of $\bar{p}$. 
Given $\mathcal{L}$ computed from $\bar{p}$, since $\mathcal{L}$ is strictly increasing, the only elements left to permute are in $\mathcal{L} \setminus \bar{p}$, which we denote with $\mathcal{D}$.
After choosing any $\sigma \in \mathcal{D}$, one applies a cyclic permutation to $\bar{p}$ that moves $\sigma$ to any position that increases the size of $\mathcal{L}$, continuing this way until the identity sequence $\iota$ is recovered.  
By sorting $\bar{p} \mapsto \iota$ by operations that (strictly) increase the size of $\mathcal{L}$, we ensure that the size of the corresponding schedule is exactly $m - \lvert \mathcal{L} \rvert$. 
  \begin{algorithm}[h!]
	\caption{Schedule construction algorithm}\label{alg:sorting}
	\setstretch{1.05}
    \begin{algorithmic}[1]
    	\Require Fixed $\bar{p} \in S_m$, LIS $\mathcal{L}$ of $\bar{p}$, and heuristic $h$
    	\Ensure $[m] = s_d \circ s_{d-1} \circ \dots \circ s_1 \circ \bar{p}$ for output sequence $\mathcal{S} = (s_i)_{i=1}^d$,
    	\Function{Schedule}{$\bar{p}$, $\mathcal{L}$, $h = \texttt{greedy}$} \Comment{See~\ref{sec:greedy} for heuristic discussion}
    		\State $(\, \mathcal{S}, \, \mathcal{D} \,  )\gets (\, \emptyset, \, [m] \setminus \mathcal{L} \,)$
    		\While{$\mathcal{D}$\text{ is not empty }}
    		    \State Select an element $k \in \mathcal{D}$ using heuristic $h$ \Comment{e.g. equation~\eqref{eq:greedy_step}}   		        	
				\State $\mathrm{k}_{\text{pred}} \gets \max \{ \ell \in \mathcal{L} \mid \ell \leq k \}$\Comment{$O(\log \log m)$ using $\mathcal{U}$}
				\State $\mathrm{k}_{\text{succ}} \gets \min \{ \ell \in \mathcal{L} \mid \ell \geq k \}$  \Comment{$O(\log \log m)$ using $\mathcal{U}$}
    		    \State $(\,i, \,i_p, \, i_n \,) \gets ( \, \bar{p}^{-1}(k), \, \bar{p}^{-1}(\mathrm{k}_{\text{pred}}), \, \bar{p}^{-1}(\mathrm{k}_{\text{succ}} ) \, )$ \Comment{$O(1)$}
				\State $j \gets \text{arbitrary } j \in [\, i_p, \, i_n)$ \textbf{if} $i < i_p$ \textbf{else} $j \in (\, i_p, \, i_n]$ \Comment{$O(1)$}
    		    \State $(\, \mathcal{S}, \mathcal{D}, \mathcal{L} \,) \gets (\, \mathcal{S} \cup (i, \, j), \, \mathcal{D} \setminus k, \, \mathcal{L} \cup k \, )$\Comment{$O(\log \log m)$}
    		    \State $\bar{p}^{-1} \gets \bar{p}^{-1} \circ m_{ij}^{-1}$ where $m_{ij}$ is given by~\eqref{eq:move_perm}\Comment{$O(m)$}
    		 \EndWhile
    		 \State \Return{$\mathcal{S}$}
    	\EndFunction
	\end{algorithmic}
\end{algorithm}

To build the schedule of permutations efficiently, we use a set-like data structure $\mathcal{U}$ that supports efficient querying the successor and predecessor of any given $s \in \bar{p}$ with respect to $\mathcal{L}$ (such as a vEB tree). 
The simplified pseudocode also uses the inverse permutation $\bar{p}$ to query the position of a given element $\sigma \in \bar{p}$, although a more efficient representation can be used via an implicit treap of the \emph{displacements} array; see Section~\ref{sec:proxy_objective}.
After $\sigma$ is inserted into $\mathcal{L}$, we update $\bar{p}$, its inverse permutations $\bar{p}^{-1}$, $\mathcal{D}$ and $\mathcal{U}$ prior to the next move. The final set of permutations to sort $\bar{p} \to \iota$ (or equivalently, $p \mapsto q$) are stored in an array $\mathcal{S}$, which is then returned for further use. 

\subsection{Greedy Counter Example}\label{sec:greedy_counter_ex}
In this section we give a simple counter-example showing that the strategy that greedily chooses valid move permutations $\{ m_{ij} \}$ minimizing the quantity
\begin{equation*}
\mathrm{cost}_{RV}(m_{ij}) = \sum\limits_{l=i+1}^j \mathds{1}\left(\, v_l(i) \neq 0 \, \right) + \sum\limits_{l=1}^m \mathds{1}\left(\, \mathrm{low}_R(l) \in [i,j] \text{ and } r_l(i) \neq 0 \, \right)
\end{equation*}
can lead to arbitrarily bad behavior. A pair of filtrations is given below, each comprising the $1$-skeleton of a $3$-simplex. Relabeling $(K, f)$ to the index set $f : K \to [m]$ and modifying $(K, f')$ accordingly yields the permutations: 
$$ (K, f) = \{ {\color{red} a \; b \; c \; d \; u \; v \; w} \; x \; y \; z  \} = { {\color{red} 1 \; 2 \; 3 \; 4 \; 5 \; 6 \; 7 } \; 8 \; 9 \; 10 } $$
$$ (K, f') = \{ {\color{red} a \; b \; c \; d } \; x \; y \; z \; {\color{red} u \; v \; w } \}  = { {\color{red} 1 \; 2 \; 3 \; 4 } \;8 \; 9 \; 10 \; {\color{red} 5 \; 6 \; 7 } }$$
The values colored in red corresponds to $\mathrm{LCS}(f, f')$. For this example, the edit distance is 
$d = m - \lvert \mathrm{LCS}(f, f') \rvert$ implies exactly $3$ moves are needed to map $f \mapsto f'$. There are six possible valid schedules of moves: 
\begin{alignat*}{3}
	S_1 = m_{x u}, m_{y u}, m_{z u} \quad & S_3 = m_{y u}, m_{x y}, m_{z u} \quad  & S_5 = m_{z u}, m_{x z}, m_{y z} \\
	S_2 = m_{x u}, m_{z u}, m_{y z} \quad  & S_4 = m_{y u}, m_{z u}, m_{x y} \quad  & S_6 = m_{z u}, m_{y z}, m_{x z} 
\end{alignat*}
where the notation $m_{x y}$ represents the move permutation that moves $x$ to the position of $y$. The cost of each move operation and each schedule is recorded in Table~\ref{table:move_costs}.
\begin{table}[h]
\caption{Move schedule costs}
\centering
\begin{tabular}{ m{0.4cm} m{0.8cm} m{0.8cm} m{0.8cm} m{0.8cm}  }
 \hline
 \multicolumn{5}{c}{Cost of each permutation} \\
 \hline
 & 1st & 2nd & 3rd & Total\\
 \hline
 $S_1$ & 2 & 3 & 1 & 6 \\
 \hline 
 $S_2$ & 2 & 2 & 4 & 8 \\
  \hline 
 $S_3$ & 4 & 2 & 2 & 8 \\
  \hline 
 $S_4$ & 4 & 3 & 3 & 10 \\
  \hline 
 $S_5$ & 2 & 2 & 4 & 8  \\
  \hline 
 $S_6$ & 2 & 5 & 3 & 10\\
 \hline
\end{tabular}
\label{table:move_costs}
\end{table}
Note the greedy strategy which always selects the cheapest move in succession would begin by moving $x$ or $z$ first, since these are the cheapest moves available, which implies one of $S_1, S_2, S_5, S_6$ would be picked on the first iteration. While the cheapest schedule $S_1$ is in this candidate set, an iterative greedy procedure would pick either $S_2$ or $S_5$, depending on the tie-breaker---thus, a greedy approach picking the lowest-cost move may not yield an optimal schedule. 
	Indeed, as the most expensive schedule $S_6$ is in initial iterations candidate set, we see that a greedy-procedure with an arbitrary tie-breaker could potentially yield a \emph{maximal-cost} schedule.  

\subsection{Crossing minimization}\label{sec:cross_minimization}
Conceptually, one way to view~\eqref{eq:upper_bound_move_cost} is as a crossing minimization problem over a set of $k-1$ bipartite graphs. To see this, consider two permutations: $p$ and $m_{ij} \circ p$, where $m_{ij}$ is a move permutation. 
Drawing $(p, m_{ij} \circ p)$ as a bipartite graph $(U, V, E)$, observe that there are $\lvert i - j \rvert$ edge crossings in the graph, and thus minimizing ~\eqref{eq:upper_bound_move_cost} is akin to a  structured variation of the $k$-layered crossing minimization problem. This is shown in an example below.
\\
\\
\textbf{Example}: Let  $p= (1\;2\;3\;4\;5\;6\;7\;8\;9)$ and $q = (9\;4\;2\;7\;1\;8\;6\;3\;5)$.
An example of three possible schedules, $S_1$, $S_2$, and $S_3$ sorting $p$ into $q$ is given in the figure below. 
\begin{figure}[!htb]
    \centering
    \includegraphics[width=0.75\textwidth]{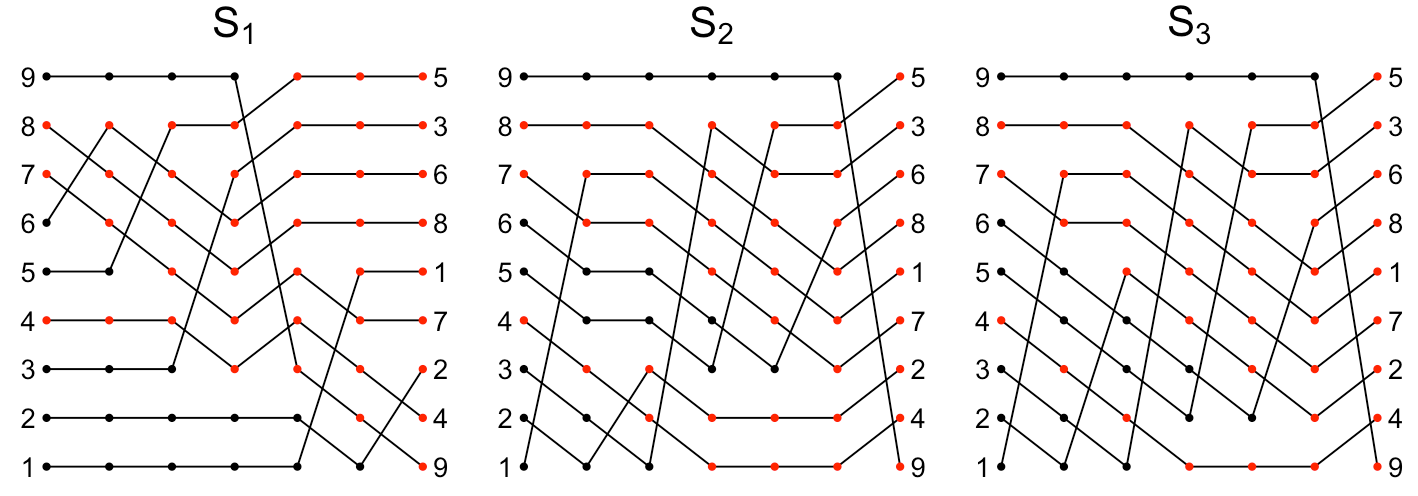} 
    \label{fig:crossings}
\end{figure}
Each column represents the successive application of a move $m_{ij}$ in the schedule, 
and the edges track the movement of each element of the permutation. 
Red vertices track the $\mathrm{LCS}$ under each permutation. 
All three schedules were generated from the same $\mathrm{LCS}(p,q) = (4\,7\,8)$ and each schedule transforms $p \mapsto q$ in $d = 6$ moves. 
In this example, $S_1$ matches the minimal number of crossings amongst all possible schedules,  since $K_\tau(p,q) = 21$. 
\\
\\
\noindent
We note that the problem we seek to solve is more structured than crossing minimization, as we are restricted to performing \emph{valid} cyclic permutations (i.e. permutations which increase the size of the LIS).

\subsection{2-parameter persistence}\label{app:2d_pers}\label{sec:fibered_barcode_reparam}
We now describe the reparameterization between the bigraded Betti numbers and the set of ``critical lines'' Lesnick and Wright~\cite{lesnick2012multidimensional} used to create their interactive 2d persistence algorithm, beginning with point-line duality.
Let $\overline{\mathcal{L}}$ denote the collection of all lines in $\mathbb{R}^2$ with non-negative slope, $\mathcal{L} \subset \overline{\mathcal{L}}$ the collection of all lines with non-negative finite slope, and $\mathcal{L}^\circ$ the collection of all affine lines with positive finite slope. 
Define the \emph{line} and \emph{point} dual transforms $\mathcal{D}_{\ell}$ and $\mathcal{D}_p$, respectively, as follows: 
\begin{equation}
	\begin{aligned}
\mathcal{D}_{\ell}: \mathcal{L} \rightarrow[0, \infty) \times \mathbb{R} & \quad \hfill \quad & \mathcal{D}_{p}:[0, \infty) \times \mathbb{R} \rightarrow \mathcal{L} \\
y=a x+b \mapsto(a,-b) &\quad \hfill \quad & (c, d) \mapsto y=c x-d
\end{aligned}
\end{equation}
The transforms $\mathcal{D}_{\ell}$ and $\mathcal{D}_p$ are \emph{dual} to each other in the sense that for any point $a \in [0, \infty) \times \mathbb{R}$ and any line $L \in \mathcal{L}$, $a \in L$ if and only if $D_\ell(L) \in D_{p}(a)$. Now, for some fixed line $L$, define  the \emph{push map} $\mathrm{push}_L(a):  \mathbb{R}^2 \to L \cup \infty$ as: 
\begin{equation}
	\mathrm{push}_L(a) \mapsto \mathrm{min}\{ v \in L \mid a \leq v \}
\end{equation}
The push map satisfies a number of useful properties. Namely: 
\begin{enumerate}
	\item For $r < s \in \mathbb{R}^2$, $\mathrm{push}_L(r) \leq \mathrm{push}_L(s)$
	\item For each $a \in \mathbb{R}^2$, $\mathrm{push}_L(a)$ is continuous on $\mathcal{L}^\circ$
	\item For $L \in \mathcal{L}^\circ$ and $S \subset \mathbb{R}^2$, $\mathrm{push}_L$ induces an ordered partition $S_L$ on $S$ 
\end{enumerate}
Property (1) elucidates how the standard partial order on $\mathbb{R}^2$ restricts to a total order on $L$ for any $L \in \overline{\mathcal{L}}$, whereas Properties (2) and (3) qualify the following definition:
\begin{definition}[Critical Lines]
	For some fixed $S \subset \mathbb{R}^2$, a line $L \in L^\circ$ is defined to be \emph{regular} if there is an open ball $B \in L^\circ$ containing $L$ such that $S_L = S_{L'}$ for all $L' \in B$. Otherwise, the line $L$ is defined as \emph{critical}. 
\end{definition}
\noindent The set of critical lines $\mathrm{crit}(M)$ with respect to some fixed set $S \subset \mathbb{R}^2$ fully characterizes a certain planar subdivision of the half plane $[0, \infty) \times \mathbb{R}$. 
This planar subdivision, denoted by $\mathcal{A}(M)$, is thus entirely determined by $S$ under point line duality.
A corollary from~\cite{lesnick2015interactive} shows that if the duals of two lines $L, L' \in \mathcal{L}$ are contained in the same $2$-cell in $\mathcal{A}(M)$, then $S_L = S_{L'}$, i.e. the partitions induced by $\mathrm{push}_L$ are equivalent. Indeed, the total order on $S_L$ is simply the pullback of the total order on $L$ with respect to the push map.
Since $\mathcal{A}(M)$ partitions the entire half-plane, the dual to every line $L \in \mathcal{L}$ is contained within $\mathcal{A}(M)$---the desired reparameterization.


To connect this construction back to persistence, one requires the definition of bigraded Betti numbers. For our purposes, the $i^{\text{th}}$-graded Betti number of $M$ is simply a function $\beta_i(M): \mathbb{R}^2 \to \mathbb{N}$ whose values indicate the number of elements at each degree in a basis of the $i^{\text{th}}$ module in a free resolution for $M$---the interested reader is referred to~\cite{lesnick2015interactive, carlsson2009theory} for a more precise algebraic definition. 
Let $S = \mathrm{supp}\,\beta_0(M) \, \cup \, \mathrm{supp}\,\beta_1(M)$, where the functions $\beta_0(M), \beta_1(M)$ are $0^{\text{th}}$ and $1^{\text{st}}$ bigraded Betti numbers of $M$, respectively. 
The main mathematical result from~\cite{lesnick2015interactive} is a characterization of the barcodes $\mathcal{B}_L(M)$, for any $L \in \mathcal{L}$, in terms of a set of \emph{barcode templates} $\mathcal{T}$ computed at every 2-cell in $\mathcal{A}(M)$.
 More formally, for any line $L \in \overline{\mathcal{L}}$ and $e$ any 2-cell in $\mathcal{A}(M)$ whose closure contains the dual of $L$ under point-line duality, the 1-parameter restriction of the persistence module $M$ induced by $L$ is given by: 
 \begin{equation}
 \mathcal{B}_L(M) = \{ [\,\mathrm{push}_L(a), \, \mathrm{push}_L(b)\,) \mid (a,b) \in \mathcal{T}^e, \mathrm{push}_L(a) < \mathrm{push}_L(b)  \} 
\end{equation}
Minor additional conditions are needed for handling completely horizontal and vertical lines. 
The importance of this theorem lies in the fact that the fibered barcodes are completely defined from the precomputed barcode templates $\mathcal{T}$---once every barcode template $\mathcal{T}^e$ has been computed and augmented onto $\mathcal{A}(M)$, $\mathcal{B}(M)$ is completely characterized, and the barcodes $\mathcal{B}_L(M)$ associated to a 1-D filtration induced by \emph{any} choice of $L$ can be efficiently computed via a point-location query on $\mathcal{A}(M)$ and a $O(\lvert \mathcal{B}_L(M) \rvert)$ application of the push map.
\\
\\
\noindent \textbf{Invariant computation: } Computationally, the algorithm from~\cite{lesnick2019computing} can be summarized into three steps: 
\begin{enumerate}
	\item Compute the bigraded Betti numbers $\beta(M)$ of $M$
	\item Construct a line arrangement $\mathcal{A}(M)$ induced by critical lines from (1) 
	\item Augment $\mathcal{A}(M)$ with \emph{barcode templates} $\mathcal{T}_e$ at every 2-cell $e \in \mathcal{A}(M)$
\end{enumerate}
Computing (1) takes approximately $\approx O(m^3)$ using a matrix algorithm similar to Algorithm~\ref{alg:reduce}~\cite{lesnick2019computing}. Constructing and storing the line arrangement $\mathcal{A}(M)$ with $n$ lines and $k$ vertices is related to the \emph{line segment intersection problem}, which known algorithms in computational geometry can solve in (optimal) output-sensitive $O((n+k) \log n)$ time~\cite{boissonnat2000robust}. 
In terms of space complexity, the number of 2-cells in $\mathcal{A}(M)$ is upper bounded by $O(\kappa^2)$, where $\kappa$ is a coarseness parameter associated with the computation of $\beta(M)$. 

There are several approaches one can use to compute $\mathcal{T}$, the simplest being to run Algorithm~\ref{alg:reduce} independently on the 1-D filtration induced by the duals of some set of points (e.g. the barycenters) lying in the interior of the 2-cells of $\mathcal{A}(M)$.  
The approach taken by~\cite{lesnick2015interactive} is to use the $R = DV$ decomposition computed at some adjacent 2-cell $e \in \mathcal{A}(M)$ to speed up the computation of an adjacent cell $e' \in \mathcal{A}(M)$. More explicitly, define the \emph{dual graph} of $\mathcal{A}(M)$ to be the undirected graph $G$ which has a vertex for every 2-cell $e \in \mathcal{A}(M)$ and an edge for each adjacent pair of cells $e, e' \in \mathcal{A}(M)$.
Each vertex in $G$ is associated with a barcode template $\mathcal{T}^e$, and the computation of $\mathcal{T}$ now reduces to computing a path $\Gamma$ on $G$ which visits each vertex at least once. To minimize the computation time, assume the $n$ edges of $G$ are endowed with non-negative weights $W = w_1, w_2, \dots, w_n$ whose values $w_i \in \mathbb{R}_{+}$ represent some notion of distance which is proportional to the computational disparity between adjacent template computations. The optimal path $\Gamma^\ast$ that minimizes the computation time is then the minimal length path with respect to $W$ which visits every vertex of $G$ at least once. There is a known $\frac{3}{2}$-approximation that can be computed efficiently which reduces the problem to the traveling salesman problem on a metric graph~\cite{christofides2022worst}, and thus can be used so long as the distance function between templates is a valid metrics. \cite{lesnick2012multidimensional} use the Kendall distance between the push-map induced filtrations, but other options are available---for example, any of the combinatorial metrics we studied in Section~\ref{sec:schedule_cost}. 

\end{document}